\documentclass[12pt,epsf]{article}
\usepackage{verbatim}
\usepackage{anyfontsize}
\usepackage{dsfont}
\usepackage{tikz}
\usepackage{sidecap}
\sidecaptionvpos{figure}{t}
\usepackage{subfigure}
\usepackage{enumitem}
\usepackage[english]{babel}
\usepackage{enumitem}  
\usepackage{amssymb,amsmath}
\usepackage{graphicx, xcolor, varwidth}
\usepackage{setspace}
\usepackage[permil]{overpic}
\usepackage{cite}
\usepackage{mathtools}
\DeclareSymbolFont{matha}{OML}{txmi}{m}{it}% txfonts
\DeclareMathSymbol{v}{\mathord}{matha}{118}

\usepackage[framemethod=default]{mdframed}
\newmdenv[skipabove=7pt,
skipbelow=7pt,
rightline=false,
leftline=false,
topline=false,
bottomline=false,
backgroundcolor=blue!10,
linecolor=blue,
innerleftmargin=5pt,
innerrightmargin=5pt,
innertopmargin=5pt,
innerbottommargin=5pt,
leftmargin=0cm,
rightmargin=0cm,
linewidth=4pt]{bBox}

\colorlet{darkblue}{blue!70!black}
\colorlet{darkgreen}{green!70!black}

\usepackage[colorlinks=true,urlcolor=darkblue,linktocpage=true,linkcolor=darkblue,citecolor=darkblue]{hyperref}

\numberwithin{equation}{section}
\DeclareMathSymbol{v}{\mathord}{matha}{118}
\newcommand{\be}{\begin{equation}}
\newcommand{\ee}{\end{equation}}
\newcommand{\bea}{\begin{eqnarray}}
\newcommand{\eea}{\end{eqnarray}}
\newcommand{\bear}{\begin{eqnarray}}
\newcommand{\eear}{\end{eqnarray}}
\newcommand{\beas}{\begin{eqnarray*}}

\newcommand{\eeas}{\end{eqnarray*}}
\newcommand{\ba}{\begin{array}}
\newcommand{\ea}{\end{array}}

\def\ba#1\ea{\begin{align}#1\end{align}}
\def\bs#1\es{\begin{split}#1\end{split}}

\newcommand{\blockS}[5]{
\begin{tikzpicture}[baseline=-3pt]
\coordinate (v1) at (-1,-0.5) {} {};
\coordinate (v2) at (-0.5,0) {} {};
\coordinate (v3) at (-1,0.5) {} {};
\coordinate (v4) at (0.5,0) {} {};
\coordinate (v5) at (1,0.5) {} {};
\coordinate (v6) at (1,-0.5) {} {};
\begin{scope}[very thick]
\draw  (v1) node[left] {$#2$}--(v2);
\draw  (v3) node[left] {$#1$}-- (v2);
\draw  (v4) -- (v2) node[midway, above] {$#5$};
\draw  (v5) node[right] {$#3$}-- (v4);
\draw  (v6) node[right] {$#4$}-- (v4);
\end{scope}
\end{tikzpicture}
}

\newcommand{\blockTT}[5]{
\begin{tikzpicture}[baseline=-3pt]
\coordinate (v1) at (-1.5,-0.5) {} {};
\coordinate (v2) at (-1,0) {} {};
\coordinate (v3) at (-1.5,0.5) {} {};
\coordinate (v4) at (1,0) {} {};
\coordinate (v5) at (1.5,0.5) {} {};
\coordinate (v6) at (1.5,-0.5) {} {};
\begin{scope}[very thick]
\draw  (v1) node[left] {$#2$}--(v2);
\draw  (v3) node[left] {$#1$}-- (v2);
\draw  (v4) -- (v2) node[midway, above] {$#5$};
\draw  (v5) node[right] {$#3$}-- (v4);
\draw  (v6) node[right] {$#4$}-- (v4);
\end{scope}
\end{tikzpicture}
}

\newcommand{\blockT}[5]{
\begin{tikzpicture}[baseline=-3pt]
\coordinate (v1) at (-0.5,1) {} {} {};
\coordinate (v2) at (0,0.5) {} {} {};
\coordinate (v3) at (0.5,1) {} {} {};
\coordinate (v4) at (0,-0.5) {} {} {};
\coordinate (v5) at (0.5,-1) {} {} {};
\coordinate (v6) at (-0.5,-1) {} {} {};
\begin{scope}[very thick]
\draw  (v1) node[left] {$#1$}--(v2);
\draw  (v3) node[right] {$#3$}-- (v2);
\draw  (v4) -- (v2) node[midway, right] {$#5$};
\draw  (v5) node[right] {$#4$}-- (v4);
\draw  (v6) node[left] {$#2$}-- (v4);
\end{scope}
\end{tikzpicture}
}

\renewcommand{\r}{\rho}
\newcommand{\br}{\bar{\rho}}

\newcommand{\pd}[2][1]{\ifnum#1=1 \frac{\partial}{\partial {#2}} \else
  \frac{\partial^#1}{\partial {#2}^{#1}}\fi}
\newcommand{\dpd}[2][1]{\ifnum#1=1 \dfrac{\partial}{\partial {#2}} \else
  \frac{\partial^#1}{\partial {#2}^{#1}}\fi}
\newcommand{\td}[2][1]{\ifnum#1=1 \frac{d}{d{#2}} \else
  \frac{d^#1}{d{#2}^{#1}}\fi}

\renewcommand{\t}{\tilde{t}}

\newcommand{\rr}{\text{\boldmath$\rho$}}

\renewcommand{\(}{\left(}
\renewcommand{\)}{\right)}

\newcommand{\nbox}{{\,\lower0.9pt\vbox{\hrule \hbox{\vrule height 0.2 cm \hskip 0.19 cm \vrule height 0.2 cm}\hrule}\,}}

\def\O{{\cal O}}

\renewcommand{\t}{\tilde{\tau}}


\begin{document}
\begin{spacing}{1.3}
\begin{titlepage}
  % end of \vbox

\begin{center}
{\Large %\bf 

A Generalized Nachtmann Theorem in CFT

}

\vspace*{6mm}

Sandipan Kundu

\vspace*{6mm}

\textit{Department of Physics and Astronomy,
\\ Johns Hopkins University,
Baltimore, Maryland, USA\\}

\vspace{6mm}

{\tt \small kundu@jhu.edu}

\vspace*{6mm}
\end{center}

\begin{abstract}
Correlators of unitary quantum field theories in Lorentzian signature obey certain analyticity and positivity properties. For  interacting unitary CFTs in more than two dimensions, we show that these properties impose general constraints on  families of minimal twist operators that appear in the OPEs of primary operators. In particular, we rederive and extend the convexity theorem which states that  for the family of minimal twist operators with even spins  appearing in the reflection-symmetric OPE of any scalar primary, twist  must be a monotonically increasing convex function of the spin. Our argument is completely non-perturbative and it also applies to the OPE of nonidentical scalar primaries in unitary CFTs, constraining the twist of spinning operators appearing in the OPE. Finally, we argue that the same methods also impose constraints on the Regge behavior of certain CFT correlators.

\end{abstract}

\end{titlepage}
\end{spacing}

\vskip 1cm
\setcounter{tocdepth}{2}  
\tableofcontents

\begin{spacing}{1.3}

%%%%%%%%%%%%%%%%%%%%%%%%%%%%%%%%%%%%%%%%%
\section{Introduction}

The operator product expansion (OPE), as introduced in \cite{Wilson:1971bg,Wilson:1971dh,Zimmermann}, provides an algebraic structure in quantum field theory (QFT) which is of fundamental importance. The OPE is particularly useful in conformal field theory (CFT) where it enjoys several nice properties. For example, two nearby {\it scalar primary} operators in any unitary CFT can be replaced by their OPE 
\be\label{ope_intro}
\O_1(x)\O_2(y)=\sum_p c_{\O_1\O_2 \O_p} D^{\mu_1 \mu_2 \cdots}_p\(x-y, \partial_y \)\O_{\mu_1 \mu_2 \cdots}^p(y)\ ,
\ee
where the sum is over primary operators which transform homogeneously under the conformal group. The coefficient function $D^{\mu_1 \mu_2 \cdots}_p$ is completely fixed by conformal symmetry up to an overall real coefficient $c_{\O_1\O_2 \O_p}$ which is commonly known as the OPE coefficient. Clearly, the operators $\O_{\mu_1 \mu_2 \cdots}^p$ that appear in the above OPE are also symmetric traceless representations of the Lorentz group and hence they can be labeled by scaling dimensions $\Delta_p$ and spins $\ell_p$. A key feature of the conformal OPE  is that after insertion into any correlator, the OPE (\ref{ope_intro})  {\it converges} absolutely even for finite separation $x-y$ \cite{Mack:1976pa,Pappadopulo:2012jk}. This fact  plays a central role, both conceptually as well as technically, in the study of CFTs. 

A CFT can be completely and uniquely defined by its OPEs. So, it is of interest to address the question of when a set of  conformal OPEs defines a unitary and causal theory. In this paper, we answer a version of this question for interacting unitary CFTs in $d>2$ dimensions by constraining the {\it family of minimal twist operators} that appears in the OPE (\ref{ope_intro}), independent of the rest of the theory. The family of minimal twist operators is defined in the following way. First, consider all spin $\ell$ primary operators that appear in the OPE (\ref{ope_intro}). Among these operators, we pick the operator $\tilde{\O}_\ell$ which has the lowest twist.\footnote{The twist is defined in the usual way $\tau=\Delta-\ell$.} We define the set of operators $\{\tilde{\O}_\ell\ |\ \ell\in {\mathbb Z}^{\ge}\}$ as the family of minimal twist operators.

Of course, any operator that appears in the OPE (\ref{ope_intro}) must obey the unitarity bound which states that in $d$ spacetime dimensions the twist $\tau_p \ge d-2$ for $\ell\ge 1$ and $\tau_p \ge \frac{d-2}{2}$ for $\ell_p=0$. Moreover, there is ample evidence suggestive of a stricter bound on the family of minimal twist operators appearing in the OPE (\ref{ope_intro}). Most notably, a concrete example of such a bound was found by Komargodski and  Zhiboedov in \cite{Komargodski:2012ek} by extending the Nachtmann theorem about  QCD sum rules of \cite{Nachtmann:1973mr} to CFT. In particular, it was argued in \cite{Komargodski:2012ek} that for a CFT to be unitary in $d>2$ dimensions, twist  must be a {\it monotonically increasing convex} function of the spin (above some lower cut-off $\ell\ge \ell_c$) for the family of minimal twist operators with even spins  appearing in any reflection-symmetric OPE $\O\O^\dagger$, where $\O$ is a scalar primary. The derivation of the convexity theorem in \cite{Komargodski:2012ek} depends on the assumptions that the CFT can flow to a gapped phase and  the {\it deep inelastic scattering} amplitude in the gapped phase is polynomially bounded in the Regge limit. However, these assumptions are not actually necessary since the convexity theorem, as shown in \cite{Costa:2017twz}, can also be derived directly using the Lorentzian inversion formula of \cite{Caron-Huot:2017vep}.  Moreover, the argument of \cite{Costa:2017twz} has the additional advantage that it implies the convexity property for the continuous even spin leading Regge trajectory above $\ell> 1$.

In this paper, we provide an alternative {\it non-perturbative} derivation of the CFT Nachtmann theorem by using analyticity and positivity properties of CFT correlators in Lorentzian signature. In fact, the  theorem that we  prove is slightly stronger than the convexity theorem of \cite{Komargodski:2012ek,Costa:2017twz}. Furthermore, our derivation can be easily generalized to constrain the family of minimal twist operators that appears in the OPE of {\it nonidentical scalar primaries} as well. Our main argument parallels the CFT-based derivation of the averaged null energy condition of \cite{Hartman:2016lgu}. It is well known that the OPE (\ref{ope_intro}) is organized as an expansion in twist in the lightcone limit. Hence, the lightcone limit of a CFT four-point function is completely fixed by the spectrum of low-twist operators. Lorentzian correlators obey certain analyticity conditions which enable us to relate the lightcone limit of a CFT four-point function to a high energy Regge-like limit. Regge correlators are theory dependent, however, they are bounded because of {\it Rindler positivity} -- which, in turn, constrains the family of minimal twist operators that contribute in the lightcone limit.

Let us now summarize our main results.  Consider the family of minimal twist operators appearing in the OPE of both $\O\O^\dagger$ and $X\overline{X}$,  where $\O$ is a scalar primary and $X$ is a completely arbitrary operator with or without spin (not necessarily local or primary) in any interacting unitary CFT in $d\ge 3$ dimensions.\footnote{Note that $\overline{X}$ is the Rindler reflection of the operator $X$. See section \ref{section_correlators} for a precise definition of Rindler reflection.} The twist $\t_\ell$ of  these minimal twist operators as a function of the spin must obey the following conditions:
\begin{itemize}
\item{\bf Monotonicity}-  For the family of minimal twist operators with even spins, twist $\t_\ell$ is a monotonically increasing function of the spin
\be\label{intro1}
\t_{\ell_2}\ge \t_{\ell_1}\ ,  \qquad \ell_2>\ell_1\ge 2\ 
\ee
for all even $\ell_1$ and $\ell_2$.
\item{\bf Global convexity}- Twist $\t_\ell$  for even spins is a non-concave function and hence  for any even $\ell_1$, $\ell_2$ and $\ell_3$
\be\label{intro2}
\frac{\t_{\ell_2}-\t_{\ell_1}}{\ell_2-\ell_1}\ge \frac{\t_{\ell_3}-\t_{\ell_1}}{\ell_3-\ell_1} \ , \qquad \ell_3>\ell_2>\ell_1\ge 2\ .
\ee
\item {\bf Local convexity}- Operators in the family of minimal twist operators with odd spins obey a local convexity condition\footnote{The OPE contains odd spin operators only when $\O$ is a complex scalar primary. }
\be\label{intro21}
\t_{\ell_o}\ge \frac{1}{2}\(\t_{\ell_o-1}+\t_{\ell_o+1}\)
\ee
for any odd $\ell_o\ge 3$.
\end{itemize}
Furthermore, the same argument also imposes nontrivial constraints on OPE coefficients of minimal twist operators. 

Conditions (\ref{intro1}) and (\ref{intro2}), in the special case $X=\O$,  are exactly the convexity theorem of \cite{Komargodski:2012ek} with $\ell_c=2$. However, the above conditions are more general. First of all, they include operators with odd spins as well. Secondly, different choices of $X$ can lead to different families of minimal twist operators and  conditions (\ref{intro1}-\ref{intro21}) apply to each such family. For example, in theories with global symmetries, the OPE of  $\O\O^\dagger$ may contain several different families of operators that obey the above conditions individually.

The preceding argument has a natural generalization  that constraints the OPE of nonidentical scalar primaries in unitary CFTs. Before we summarize these constraints for two arbitrary  nonidentical scalar primaries $\O_1$ and $\O_2$, we should introduce some basic notations.  First, consider all spin $\ell$ operators that appear in the OPE $\O_1\O_2$. Among these operators, we pick the lowest twist operator $\tilde{\O}^{(12)}_\ell$ which has twist $\t^{(12)}_\ell$. The set of operators $\{\tilde{\O}^{(12)}_\ell\ |\ \ell\in {\mathbb Z}^{\ge}\}$ is defined as the the family of minimal twist operators in the OPE $\O_1\O_2$. Similarly, we can define families of minimal twist operators for the OPEs $\O_1 \O_1^\dagger$ and  $\O_2 \O_2^\dagger$. Twists of these operators are denoted by  $\t^{(11)}_\ell$ and $\t^{(22)}_\ell$ respectively. A Nachtmann-like theorem for nonidentical scalar primaries relates these three families of minimal twist operators. In particular,  the twist $\t^{(12)}_\ell$ as a function of the spin must obey the following conditions in any interacting unitary CFT in $d\ge 3$ dimensions.
\begin{itemize}
\item{\bf Lower bound for even spins}- The family of minimal twist operators that appears in the OPE $\O_1\O_2$  must obey a lower bound for even spins
\be
\tilde{\tau}^{(12)}_{\ell_e}\ge \frac{1}{2}\(\tilde{\tau}^{(11)}_{\ell_e}+\tilde{\tau}^{(22)}_{\ell_e} \)\ , \qquad \text{for even}\ \ell_e\ge 2\ . 
\ee
\item{\bf Lower bound for odd spins}- The family of minimal twist operators that appears in the OPE $\O_1\O_2$  must obey a different lower bound for odd spins
\be\label{intro5}
\tilde{\tau}^{(12)}_{\ell_o}\ge \frac{1}{2}\(\frac{\tilde{\tau}^{(11)}_{\ell_o-1}+\tilde{\tau}^{(22)}_{\ell_o-1}}{2}+\frac{\tilde{\tau}^{(11)}_{\ell_o+1}+\tilde{\tau}^{(22)}_{\ell_o+1} }{2}\)\ , \qquad \text{for odd}\ \ell_o\ge 3\ .
\ee
\end{itemize}
Thus for a CFT to be unitary and causal in $d\ge 3$ dimensions, twist  for the family of minimal twist operators with even (or odd) spins  appearing in the OPE $\O_1\O_2$ must be bounded from below by a  monotonically increasing convex function of the spin. Moreover, crossing implies that for asymptotically large spins $\t_\ell \rightarrow \Delta_1+\Delta_2$, where $\Delta_1$ and $\Delta_2$ are dimensions of $\O_1$ and $\O_2$ respectively \cite{Komargodski:2012ek, Fitzpatrick:2012yx}.

In recent years, the conformal bootstrap has made significant progress in constraining CFT data, both numerically and analytically, by using crossing symmetry, unitarity, and conformal symmetry. We closely examine some of the bootstrap results in $d\ge 3$ dimensions to demonstrate that OPEs in unitary CFTs  indeed obey the above conditions. In particular, the numerical results of \cite{Simmons-Duffin:2016wlq} enable us to study and compare families of minimal twist operators that appear in the OPEs $\sigma\sigma$, $\epsilon\epsilon$, and $\sigma\epsilon$ of the 3d Ising CFT, where operators $\sigma$ and $\epsilon$ are the lowest-dimension $\mathbb{Z}_2$-odd and $\mathbb{Z}_2$-even scalar primaries respectively (see figure \ref{ising}). Furthermore, we show that the lightcone bootstrap results of anomalous dimensions of various double-twist operators  at large spins are  consistent with the conditions (\ref{intro1})-(\ref{intro5}). Of course, our non-perturbative results are also in complete agreement with systematic large spin perturbation theory results of \cite{Alday:2015eya,Alday:2015ota,Alday:2016njk}.

Finally, we argue that the same methods can be deployed in the Regge limit to constrain the Regge behavior of certain CFT correlators. However, these constraints are expected to be theory dependent since the Regge limit is dominated by high spin exchanges which are in general non-universal. Nevertheless, this approach is still useful, particularly in the context of the AdS/CFT correspondence, where the constraints on the Regge correlator should be interpreted as bounds on various interactions of low energy effective field theories in AdS. 

The outline of this paper is as follows.  In section \ref{section_correlators} we review analyticity and positivity properties of QFT correlators in Lorentzian signature. In section \ref{section_cft} we utilize these properties to derive a stronger version of the CFT Nachtmann theorem in unitary CFTs in $d\ge3$ dimensions. We then extend our analysis in section \ref{section_mixed} to impose constraints on families of minimal twist operators that appear in the OPE of nonidentical scalar primaries. All these constraints are consistent with the conformal bootstrap results, as we discuss in section \ref{section_ising}. Finally in section \ref{section_regge} we explain how the same methods could be useful in constraining the Regge behavior of certain CFT correlators.

\section{Analyticity and Positivity of Lorentzian Correlators} \label{section_correlators}
A QFT can be uniquely defined by its Euclidean correlators. Lorentzian correlators of any ordering can then be obtained from Euclidean correlators by performing appropriate analytic continuations. The Osterwalder-Schrader reconstruction theorem guarantees that well behaved Euclidean correlators, upon analytic continuation, produce Lorentzian correlators that
obey the Wightman axioms \cite{Osterwalder:1973dx,Osterwalder:1974tc}.\footnote{Recently, it has been pointed out that the reconstruction theorem has limitations because of certain technical caveats.  Nevertheless, the relation between well behaved Euclidean correlators and well behaved Lorentzian correlators can be made precise in CFT \cite{slava1,slava2}.}  Furthermore, the analytic structure of Lorentzian correlators of unitary QFTs follows from the fact that Euclidean correlators are single-valued, permutation invariant functions of positions that do not have any branch cuts as long all points remain Euclidean \cite{haag}. In particular, in $d$ spacetime dimensions any Lorentzian correlator of operators with or without spins
\be
\langle O_1(x_1)O_2(x_2)O_3(x_3) \cdots\rangle \nonumber
\ee
as a function of complexified $x_i$ is analytic in the domain \cite{haag,Hartman:2016lgu}
\be\label{con1}
\text{Im}\ x_1 \vartriangleleft \text{Im}\ x_2  \vartriangleleft \text{Im}\ x_3  \vartriangleleft \cdots\ ,
\ee
where, $\mbox{Re}\ x_i \in \mathbb{R}^{1,d-1}$ and $\mbox{Im}\ x_i \in \mathbb{R}^{1,d-1}$. The symbol $x \vartriangleleft y$ represents that the point $x$ is in the past lightcone of point $y$. This analyticity condition is simply a covariant version of the standard $i\epsilon$ prescription that computes Lorentzian correlators from Euclidean correlators.\footnote{The full domain of analyticity of Lorentzian correlators, as described in \cite{haag}, is even larger than (\ref{con1}). However, for the purpose of this paper, the condition (\ref{con1}) is sufficient. } Still, analyticity of Lorenzian correlators in the domain (\ref{con1}) is a nontrivial statement since it is intimately related to microcausality in Lorentzian signature  \cite{Hartman:2016lgu}. 

Well behaved Euclidean correlators must also satisfy \textit{reflection positivity} which is closely related to unitarity. In Lorentzian signature, there is a positivity property known as Rindler positivity which is physically very similar to reflection positivity \cite{Casini:2010bf,Maldacena:2015waa}.\footnote{See \cite{Hartman:2016lgu} for a review and \cite{Rosso:2019txh} for a generalization to arbitrary CFTs in the Lorentzian cylinder and in de Sitter.} It states that all Rindler reflection symmetric correlators must be positive in any unitary, Lorentz-invariant QFT. In order to be more specific, let us introduce the following notation for points $x \in \mathbb{R}^{1,d-1}$:
\be
x = (t,y,\vec{x})\equiv (x^-,x^+,\vec{x})\ , 
\ee
where, $x^-=t-y$ and $x^+=t+y$. The right and left Rindler wedges are defined as $R=\{(x^-,x^+,\vec{x}): x^+>0,x^-<0\}$, $L=\{(x^-,x^+,\vec{x}): x^+<0,x^->0\}$. The Rindler reflection takes a point on $R$ to a point on $L$ and vice versa. In general, the Rindler reflection of an operator is defined as 
\be
\overline{O_{\mu_1 \mu_2 \cdots}(t,y,\vec{x})}=(-1)^P O^{\dagger}_{\mu_1 \mu_2 \cdots}(-t^*,-y^*,\vec{x})\ ,
\ee
where, $P$ is the total number of $t$ and $y$ indices. Rindler positivity is the statement that the Lorentzian correlator 
\be\label{con2}
\langle  O_1(x_1) \cdots O_n(x_n) \overline{O_1(x_1)} \cdots \overline{O_n(x_n)}\rangle >0 \ ,
\ee
where all $x_i \in L$ (or equivalently in $R$) and operators inside the correlator is ordered as written. 

These analyticity and positivity conditions hold for any unitary Lorentzian QFT making them important tools that can be employed to derive very general results. In particular, these simple properties have far-reaching  consequences for theories with conformal symmetry as we discuss next. 

\section{CFT in Lorentzian Signature} \label{section_cft}

We now consider unitary CFTs in $d>2$ spacetime dimensions. First we show that analyticity and positivity properties of Lorentzian correlators imply the CFT Nachtmann theorem above $\ell\ge 2$.

We start with Lorentzian correlators
\be\label{corr}
G=\frac{\langle O_2(\mathbf{1}) O_1(\rr)\overline{O_1(\rr)}\ \overline{O_2(\mathbf{1})} \rangle}{ \langle O_2(\mathbf{1}) \overline{O_2(\mathbf{1}) }\rangle \langle  O_1(\rr)\overline{O_1(\rr) }\rangle} \, \qquad G_0=\frac{\langle O_2(\mathbf{1})   O_1(\rr)\overline{O_2(\mathbf{1}) }\ \overline{O_1(\rr) }\rangle}{\langle O_2(\mathbf{1}) \overline{O_2(\mathbf{1}) }\rangle \langle  O_1(\rr)\overline{O_1(\rr) }\rangle}
\ee
of two arbitrary CFT operators $O_1$ and $O_2$ (with or without spin), where operators inside correlators are ordered as written. All the points are restricted to be in a 2d subspace   
\be\label{points}
\mathbf{1}\equiv(t=0,y=-1,\vec{0})\ , \qquad \rr\equiv(x^-=\r,x^+=-\br,\vec{0})\ 
\ee
with $1>\br>0$ and $\r>1$, as shown in figure \ref{config}.  Notice that operator pairs $O_2(\mathbf{1}), O_1(\rr)$ and $\overline{O_1(\rr)}, \overline{O_2(\mathbf{1})}$ are time-like separated. Clearly, the correlator $G$ in (\ref{corr}) is not Rindler reflection symmetric, however, it is related to the Rindler reflection symmetric correlator $G_0$ by an analytic continuation. 

\begin{figure}
\begin{center}
\begin{tikzpicture}[baseline=-3pt, scale=1.80]
\begin{scope}[very thick,shift={(4,0)}]
\coordinate (v1) at (-1.5,-1.5) {};
\coordinate(v2) at (1.5,1.5) {};
\coordinate (v3) at (1.5,-1.5) {};
\coordinate(v4) at (-1.5,1.5) {};

\draw[thin,-latex]  (v1) -- (v2)node[left]{$x^+$};
\draw[thin,-latex]  (v3) -- (v4)node[right]{$\ x^-$};
\draw(-1.5,0)node[left]{ $\ O_2(\mathbf{1})$};
\draw(1.5,0)node[right]{ $ \overline{O_2(\mathbf{1})}$};
\filldraw[black]  (-1.5,0) circle (1pt);
\filldraw[black]  (1.5,0) circle (1pt);
\coordinate(v5) at (0,0) {};
\def \fac {.6};
\filldraw[black]  (-1.2,1) circle (1 pt);
\filldraw[black]  (1.2,-1) circle (1pt);
\draw(-1.2,1)node[left]{ $ O_1(\rr)$};
\draw(1.2,-1)node[right]{ $ \overline{O_1(\rr)}$};

\end{scope}
\end{tikzpicture}
\end{center}
\caption{\label{config} \small A Lorentzian four-point function where all points are restricted to a $2$d subspace. Null coordinates are defined as $x^{\pm}=t\pm y$, where time is running upward.}
\end{figure}

Analyticity and positivity properties of Lorentzian correlators, as discussed in  \cite{Hartman:2016lgu}, impose nontrivial constraints on $G$.  These constraints can be conveniently described by introducing
\be\label{sigma}
\r=\frac{1}{\sigma}\ , \qquad \br=\sigma \eta  
\ee
with $\eta>0$ which makes $G\equiv G(\eta,\sigma)$ and $G_0\equiv G_0(\eta,\sigma)$. In any unitary QFT, correlators $G(\eta,\sigma)$ and $G_0(\eta,\sigma)$ in the regime $0<\eta<1$ have the following properties:
\begin{enumerate}[label=(\roman*)]
\item \label{analytic}  The analyticity condition (\ref{con1}) implies that $G(\eta,\sigma)$, as a function of complex $\sigma$, is analytic near $\sigma \sim 0$ on the lower-half $\sigma$-plane.
\item \label{bound} Rindler positivity naturally defines a positive inner product which comes equipped with a Cauchy-Schwarz inequality. The Cauchy-Schwarz inequality implies that the real part of the correlator $G(\eta,\sigma)$ for real $\sigma$ with $|\sigma| < 1$ is bounded \cite{Hartman:2016lgu}
\be\label{con3}
\mbox{Re}\ G(\eta,\sigma) \leq  G_0(\eta,\sigma)\ ,
\ee
where Rindler positivity also requires that $G_0(\eta,\sigma)>0$.
\end{enumerate}
Note that both these conditions are valid even when $O_1$ and $O_2$ are smeared.

Conditions \ref{analytic} and \ref{bound} together lead to nontrivial bounds on the spectrum of low-lying operators in interacting CFTs in $d>2$ dimensions. The main argument is  actually very intuitive. It is well known that CFT correlators in the \textit{lightcone limit} $\eta\ll |\sigma|\ll 1$  have universal features because these correlators are completely fixed by the spectrum of low-twist operators. Now, the condition \ref{analytic} which is basically causality in disguise relates the lightcone limit to a high energy Regge-like limit $|\sigma|\ll 1$. In general, we do not have computational control on CFT correlators in the limit $|\sigma|\ll 1$, however, these correlators are bounded because of the ``unitarity" condition \ref{bound} -- which, in turn, imposes constraints on the spectrum of low-twist operators that contribute in the lightcone limit. 

\subsection{Lightcone Limit} 
\begin{figure}
\centering
\includegraphics[scale=0.5]{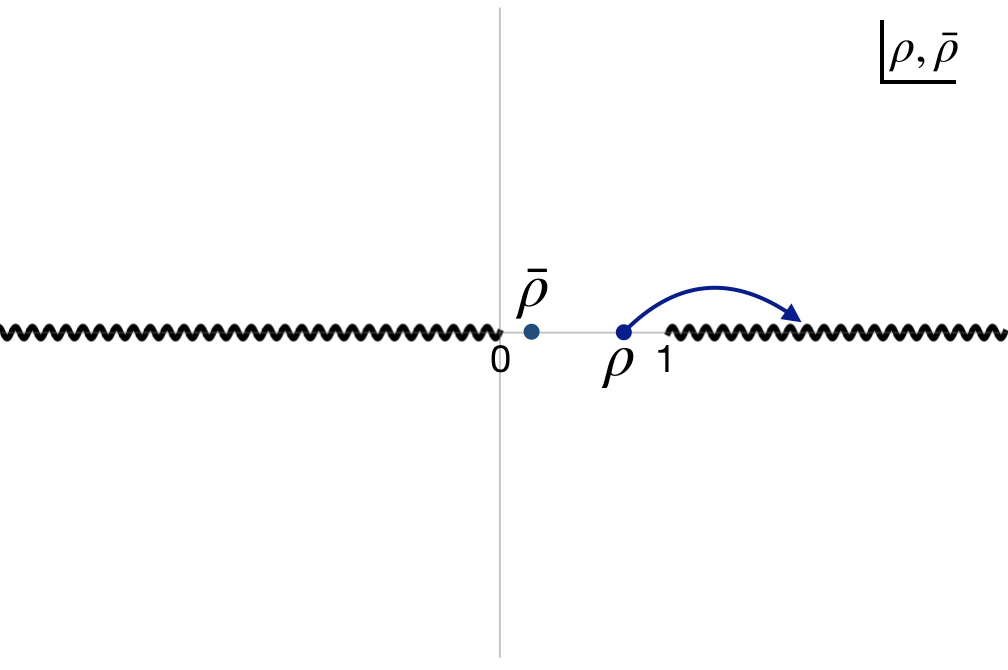}
\caption{ \label{dt1} \small The Lorentzian correlator $G$ of (\ref{corr}) is obtained from the Euclidean correlator by analytically continuing $\r$ along the blue path.}
\end{figure}

We now restrict to a CFT correlator where $O_1=\O$ is a \textit{scalar primary} (real or complex) and $O_2=X$ is some arbitrary operator with or without spin (not necessarily local or primary). We are interested in  the lightcone limit 
\be\label{lightcone}
\r\br \rightarrow 0\ , \qquad \text{then} \qquad \r\rightarrow \infty
\ee
of the correlator $\langle X(\mathbf{1}) \O(\rr)\overline{\O(\rr)}\ \overline{X(\mathbf{1})} \rangle$. This can be achieved by generalizing the lightcone OPE of \cite{Hartman:2016lgu} for complex  scalar primaries. We start with the OPE 
\be\label{ope}
\O(x)\O^\dagger(y)= \sum_p c_{\O\O^\dagger \O_p} D^{\mu_1 \mu_2 \cdots}_p\(x-y, \partial_y \)\O_{\mu_1 \mu_2 \cdots}^p(y)
\ee
where the sum is only over primary operators that have non-vanishing three-point functions with $\O\O^\dagger$. The contribution of the primary operator $\O^p_{\mu_1\mu_2\cdots}$ with conformal dimension $\Delta_p$ and spin $\ell_p$ and its descendants to the OPE $\O\O^\dagger$ in the lightcone limit $\r\br \rightarrow 0$ can be recast as 
\begin{align}\label{lope}
\frac{\O(\rr)\O^\dagger(-\rr)|_{\O_p} }{\langle \O(\rr)\O^\dagger(-\rr)\rangle}=\lambda_p (\r\br)^{\frac{\tau_p}{2} }\r^{\ell_p-1}  \int_{-\r}^\r d\r' \left(1-\frac{{\r'}^2}{\r^2} \right)^{\frac{\Delta_{p}+\ell_p}{2}-1} \O^p_{--\cdots -}(\r',0,\vec{0}) 
\end{align}
where, twist $\tau_p=\Delta_p-\ell_p$ and $\lambda_p$ is a numerical coefficient exact value of which will not be important for us.\footnote{The exact value of $\lambda_p$ can be computed from the three-point function $\langle \O\O^\dagger\O_{\mu_1 \mu_2 \cdots}^p\rangle$ 
  \be\nonumber
  \lambda_p=\frac{(-1)^{\ell_p}    c_{\O\O^\dagger\O_p} 2^{\Delta_p} \Gamma \left(\frac{\Delta_p+\ell_p+1}{2}\right)}{c_p \sqrt{\pi}\Gamma \left(\frac{\Delta_p+\ell_p}{2} \right)^2}\ ,
  \ee
  where,  $c_p$ is the coefficient of the two-point function of $\O_{\mu_1 \mu_2 \cdots}^p$. Note that for real scalar primaries only operators with even spins can appear in the OPE. Hence, the factor of $(-1)^{\ell_p}$ is not there in \cite{Hartman:2016lgu} for real scalars.} This lightcone OPE can be used inside arbitrary correlators.  In particular, we can use this OPE to derive the contribution of the $\O\O^\dagger\rightarrow\O_p\rightarrow X\overline{X}$ conformal block to the correlator $\langle X(\mathbf{1}) \O(\rr)\overline{\O(\rr)}\ \overline{X(\mathbf{1})} \rangle$ in the Lorentzian lightcone limit (\ref{lightcone}). This Lorentzian correlator can be obtained from the Euclidean one by analytically continuing $\r$ along the path shown in figure \ref{dt1}.  The corresponding contour for the $\r'$-integral  follows directly from the way operators are ordered inside the correlator.\footnote{See \cite{Hartman:2016lgu} for a detailed discussion.}
 
 Of course, we intend to utilize the analyticity and positivity conditions \ref{analytic} and \ref{bound}. Hence, it is more convenient to study the lightcone limit of normalized correlators $G(\eta,\sigma)$ and $G_0(\eta,\sigma)$, as defined in (\ref{corr}),  using variables (\ref{sigma}). In these variables, the Lorentzian lightcone limit (\ref{lightcone}) can be equivalently defined as  $0<\eta\ll |\sigma|\ll 1$. As an immediate consequence of the lightcone OPE (\ref{lope}), we can write the contribution of the operator $\O_p$ to $G(\eta,\sigma)-G_0(\eta,\sigma)$ in the Lorentzian lightcone limit as an expansion 
\begin{align}\label{block}
G(\eta,\sigma)|_{\O_p}-G_0(\eta,\sigma)|_{\O_p}=i\frac{\eta^{\frac{\tau_p}{2}}}{\sigma^{\ell_p-1}}\left(C_{\ell_p,\ell_p} +\sum_{n=2,4,\cdots}C_{\ell_p,\ell_p-n}\sigma^n\right) +\cdots\ ,
\end{align}
where dots represent correction terms that are either suppressed by higher powers of $\eta$ or decay for small $\sigma$. Notice that the sum in the above equation is only over even integers. Moreover, coefficients $C_{\ell_p,\ell'}$ are real which can be computed from the lightcone OPE (\ref{lope}). For example
\be\label{hsanec}
C_{\ell_p,\ell_p}= \lambda_p \frac{\mbox{Im}\ \langle X(\mathbf{1})\int_{-\infty}^\infty d\r' \O^p_{--\cdots -}(\r',0,\vec{0})  \overline{X(\mathbf{1})}\rangle}{\langle X(\mathbf{1}) \overline{X(\mathbf{1})}\rangle}
\ee
which is nonzero by definition and all other coefficients $C_{\ell_p,\ell'}\propto C_{\ell_p,\ell_p}$. Since, however, these coefficients do not depend on $\eta$ or $\sigma$, their actual values are not important to us.  

By using  the $i\epsilon$ prescription we can convince ourselves that that we do not need to cross any branch cuts to obtain a Rindler reflection symmetric correlator, even when some of the operators are time-like separated. Hence, the correlator $G_0(\eta,\sigma)$ can be computed by using the Euclidean OPE. This implies that $G_0(\eta,\sigma)$ is trivial in the lightcone limit (\ref{lightcone}) where it is dominated by the identity exchange
\be\label{g0approx}
G_0(\eta,\sigma) = 1+\cdots\ , 
\ee
where dots represent correction terms that are suppressed by positive powers of $\eta$ and $\sigma$.

The $\O\O^\dagger$ OPE converges only when all operators are spacelike separated in a correlator. On the other hand, the $\O\O^\dagger$ OPE in a Lorentzian correlator such as $G(\eta,\sigma)$ may diverge. Notice that  contributions of individual operators in (\ref{block}) become increasingly singular with increasing spin which is a manifestation of the fact that the lightcone OPE (\ref{lope}) does not converge on the second sheet. However, by considering scalar four-point functions  it was argued in \cite{Hartman:2015lfa} that the second sheet conformal blocks still can be trusted in the lightcone limit. The same argument applies in general implying any finite number of terms in the  lightcone OPE on the second sheet  produce a reliable asymptotic expansion in the limit $\eta\rightarrow 0$ \cite{Hartman:2016lgu}. Hence, $G(\eta,\sigma)$ in the Lorentzian lightcone limit  can be expressed as a divergent asymptotic series which is organized by twist\footnote{In the equation (\ref{sinp}), we have ignored terms that decay for small $\sigma$. Any term in $G(\eta,\sigma)$ that decay for small $\sigma$ can be subtracted by analytically continuing $G_0(\eta,\sigma)$ appropriately and hence these terms do not play any role in our argument. This has been explained in detail in appendix \ref{appendixA}.  }
\be\label{sinp}
G(\eta,\sigma) \sim G_0(\eta,\sigma) +i\sum_{\tau_p\le \tau_{cutoff}} \frac{\eta^{\frac{\tau_p}{2}}}{\sigma^{\ell_p-1}}\left(C_{\ell_p,\ell_p} +\sum_{n=2,4,\cdots}C_{\ell_p,\ell_p-n}\sigma^n+\cdots\right) \ ,
\ee
where, dots represent terms that are suppressed by positive powers of both $\eta$ and $\sigma$. The spectrum of operators that appear both in the OPE $\O\O^\dagger$ and $X\overline{X}$ can have an accumulation point in the twist. In that case, it is generally expected that the above asymptotic expansion is not valid  beyond the accumulation point in the twist spectrum.

\subsection{Nachtmann Theorem in CFT} \label{nachtmann}
We now have all the ingredients to derive and generalize the CFT Nachtmann theorem. Our starting point is the correlator $\langle X(\mathbf{1}) \O(\rr)\overline{\O(\rr)}\ \overline{X(\mathbf{1})} \rangle$, where $\O$ is a scalar primary and $X$ is a completely arbitrary operator.  The lightcone limit of this conformal four-point function, as we learned from equation (\ref{sinp}),  is an expansion in twist. So, in this limit only the family of minimal twist operators is important. 

The family of minimal twist operators is defined in the usual way. In the present context, consider all spin $\ell$ operators that appear in the OPE of both $\O\O^\dagger$ and $X \overline{X}$. Among these operators, we pick the operator $\tilde{\O}_\ell$ which has the lowest twist.  The family of minimal twist operators is then defined as $\{\tilde{\O}_\ell\ |\ \ell\in {\mathbb Z}^{\ge}\}$. We adopt the notation $\t_\ell$ to denote the twist of $\tilde{\O}_\ell$. 

Clearly, the spectrum of operators that appear in the OPE of both $\O\O^\dagger$ and $X \overline{X}$ can have both even and odd spins when $\O$ is a complex scalar. In that case, it is more convenient to discuss the even and odd (spin) family of minimal twist operators separately.

\begin{figure}[h]
\begin{center}
\includegraphics[width=0.55\textwidth]{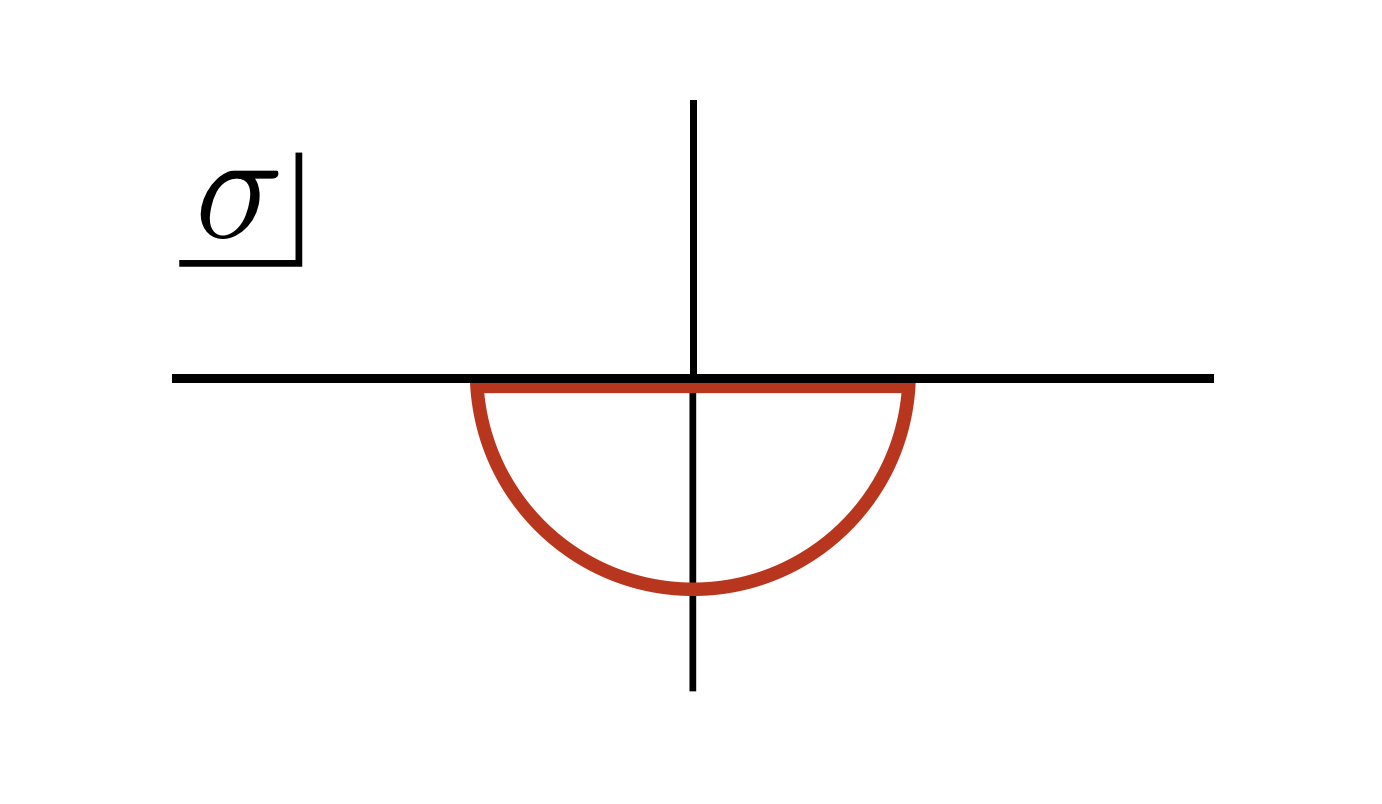}
\caption{The integration contour in the complex-$\sigma$ plane.}\label{contour}
\end{center}
\end{figure}

First, we  derive a general sum rule by integrating $\delta G(\eta,\sigma)$, where
 \be\label{deltaG}
 \delta G(\eta,\sigma)\equiv G_0(\eta,\sigma)- G(\eta,\sigma)\ .
 \ee
 As in \cite{Hartman:2016lgu}, we choose an integration contour which is   a half-disk near $\sigma=0$, just below the real line in the complex-$\sigma$ plane  (see figure \ref{contour}). The radius $R$ of the  semicircle is such that it satisfies $\eta\ll R\ll 1$. Strictly speaking, $ G_0(\eta,\sigma)$ is only defined on the real line. However, we can still define a function $G_0^{(- )}(\eta,\sigma)$ which is analytic on the lower half $\sigma$-plane and has the property $\mbox{Re}\ G_0^{(- )}(\eta,\sigma) = G_0(\eta,\sigma)$ (for a detailed discussion see appendix \ref{appendixA}). The  analyticity property \ref{analytic} then requires that 
\be
\mbox{Re} \oint d\sigma\ \sigma^n \( G_0^{(- )}(\eta,\sigma) -G(\eta,\sigma)\) =0 \ 
\ee
for integer $n\ge0$. The above contour integral can be divided into two parts -- one over the real line and one over the semicircle. The lightcone limit expression (\ref{block}) is valid over the entire semicircle. Moreover, the semicircle integral can be further and crucially simplified using the identity
\be\label{sachin}
\mbox{Re}\ i \int_S d\sigma \sigma^n = \pi \delta_{n,-1}\ ,
\ee
where, $S=\{R e^{i\theta}, \theta\in [0,-\pi]\}$. This enables us to write a sum rule
\be\label{sum}
\sum_{\ell'\ge \ell}C_{\ell', \ell}\ \eta^{\frac{\t_{\ell'}}{2}}=\frac{1}{\pi}\int_{-R}^R d\sigma\ \sigma^{\ell-2} \mbox{Re}(\delta G(\eta,\sigma))\ , \qquad \text{integer}\ \ell\ge 2
\ee
 where $0<\eta\ll R\ll 1$. The left hand side of the above relation is written as a formal sum which reflects the fact that any operator $\tilde{\O}_{\ell'}$ with $\ell'=\ell+2N$ (for non-negative integer $N$) can contribute to the correlator $G(\eta,\sigma)$  a term that grows exactly as $\frac{1}{\sigma^{\ell-1}}$ in the Lorentzian lightcone limit. However, it should be understood that for any $\ell$ only the term (or terms) with the lowest $\t_{\ell'}$ dominates.  

The observant reader may have noticed that dots in  equation (\ref{block}) contain terms that decay with  positive but non-integer powers of $\sigma$ when operators with non-integer dimensions are exchanged. These terms appear to spoil the projection to a specific power of $\frac{1}{\sigma}$ by using the identity (\ref{sachin}). However, these terms do not actually contribute to the sum rule (\ref{sum}), since these terms get exactly canceled by $G_0^{(- )}(\eta,\sigma)$. This is explained in detail in  appendix \ref{appendixA}.

Next we use the sum rule (\ref{sum}) to derive the monotonicity and convexity properties of the even spin family of minimal twist operators. 

\subsubsection*{Monotonicity}
The positivity condition \ref{bound} immediately implies that the left hand side of (\ref{sum}) must be non-negative for any even $\ell$.\footnote{Strictly speaking, the inequality (\ref{con3}) is valid only for real $\sigma$. There is a  complexified version of Rindler positivity, as discussed in \cite{Hartman:2016lgu}, which implies (\ref{con3}) on the straight line part of the contour \ref{contour} even when it lies just below the real axis in the complex-$\sigma$ plane.} Furthermore, the positivity condition \ref{bound} also leads to the following inequality for any two even $\ell_2>\ell_1\ge 2$
\be\label{proof1}
\frac{\int_{-R}^R d\sigma\ \sigma^{\ell_2-2} \mbox{Re}(\delta G(\eta,\sigma))}{\int_{-R}^R d\sigma\ \sigma^{\ell_1-2} \mbox{Re}(\delta G(\eta,\sigma))}\le R^{\ell_2-\ell_1}\ll 1\ .
\ee
Hence, the sum rule (\ref{sum}) implies that
\be\nonumber
\frac{\sum_{\ell'\ge \ell_2}C_{\ell', \ell_2}\ \eta^{\frac{\t_{\ell'}}{2}}}{\sum_{\ell'\ge \ell_1}C_{\ell', \ell_1}\ \eta^{\frac{\t_{\ell'}}{2}}}\ll 1
\ee
in the limit $\eta\rightarrow 0$. Since, $C_{\ell,\ell'}$ do not scale with $\eta$, the above inequality can be satisfied  iff 
\be\label{mono}
\t_{\ell_2}\ge \t_{\ell_1}\ ,  \qquad \ell_2>\ell_1\ge 2\ 
\ee
for all even $\ell_1$ and $\ell_2$. The equality in the above relation holds only for free CFTs and 2d CFTs. In these CFTs, the relation (\ref{mono}) is trivially satisfied since they contain an infinite tower of higher spin conserved currents.   

The monotonicity condition (\ref{mono}) greatly simplifies the sum rule (\ref{sum}) for interacting CFTs in $d>2$ dimensions. For any even $\ell$, clearly the first term $C_{\ell,\ell}\eta^{\frac{\t_\ell}{2}}$ dominates in the left hand side implying three-point functions
\be
C_{\ell,\ell}> 0 \qquad \text{for even $\ell\ge 2$}
\ee
which is consistent with the averaged null energy condition and its higher spin generalization as derived in \cite{Hartman:2016lgu}.

\subsubsection*{Convexity}
Similar to \cite{Komargodski:2012ek}, we can actually derive a stronger constraint on the family of minimal twist operators with even spins. For any even $\ell\ge2$, the Cauchy-Schwarz inequality of real integrable functions imposes
\be\label{proof2}
0< \frac{\(\int_{-R}^R d\sigma\ \sigma^{\ell} \mbox{Re}(\delta G(\eta,\sigma))\)^2}{\int_{-R}^R d\sigma\ \sigma^{\ell-2} \mbox{Re}(\delta G(\eta,\sigma)) \int_{-R}^R d\sigma\ \sigma^{\ell+2} \mbox{Re}(\delta G(\eta,\sigma)) }\le 1\ .
\ee
In the regime $0<\eta\ll R\ll 1$,  we can use the sum rule (\ref{sum}) and the monotonicity condition (\ref{mono}) to rewrite the above inequality as
\be
0< \lim_{\eta\rightarrow 0} \(\frac{C_{\ell+2,\ell+2}^2}{C_{\ell,\ell}C_{\ell+4,\ell+4}} \)\eta^{\frac{1}{2}\(2\t_{\ell+2}-\t_{\ell}-\t_{\ell+4}\)}\le 1
\ee
implying $2\t_{\ell+2}\ge \t_{\ell}+\t_{\ell+4}$ for any even $\ell\ge 2$. This local convexity condition can  be applied iteratively to obtain 
\be\label{convex}
\frac{\t_{\ell_3}-\t_{\ell_1}}{\ell_3-\ell_1}\le \frac{\t_{\ell_2}-\t_{\ell_1}}{\ell_2-\ell_1}\ , \qquad 2\le \ell_1<\ell_2<\ell_3
\ee
for all even $\ell_1$, $\ell_2$ and $\ell_3$.

To summarize, twist is a monotonically increasing convex function of the spin for the even (spin) family of minimal twist operators appearing in the OPE of both $\O\O^\dagger$ and $X\overline{X}$,  where $\O$ is a scalar primary and $X$ is a completely arbitrary operator with or without spin (not necessarily local or primary) in any interacting unitary CFT in $d\ge 3$ dimensions. This is, in general, stronger than the convexity theorem of \cite{Komargodski:2012ek} (with $\ell_c=2$). In particular, different choices of $X$ can lead to different families of minimal twist operators and the conditions (\ref{mono}) and (\ref{convex}) apply to each such family. For example, in theories with global symmetries, the OPE   $\O\O^\dagger$ may contain several different families of operators that obey the monotonicity and the convexity conditions.

\subsubsection*{Family of minimal twist operators with odd spins}
Let us now move on to minimal twist operators with odd spins. This discussion is relevant only when $\O$ is complex. Clearly, the integrand in (\ref{sum}) is not strictly positive for odd $\ell$. Hence, the family of minimal twist operators with odd spins does not in general obey the monotonicity and the convexity conditions. This should not be surprising since  there are known examples that would violate any such conditions \cite{Li:2015rfa}. However, it is not true that the family of minimal twist operators with odd spins is completely free of constraints. For any even $\ell_e$ and odd $\ell_o$ satisfying $\ell_o>\ell_e\ge2$, the positivity condition \ref{bound} imposes
\be\label{proof22}
\frac{|\sum_{\ell'\ge \ell_o}C_{\ell', \ell_o}\ \eta^{\frac{\t_{\ell'}}{2}}|}{C_{\ell_e, \ell_e}\ \eta^{\frac{\t_{\ell_e}}{2}}}
=\frac{|\int_{-R}^R d\sigma\ \sigma^{\ell_o-2} \mbox{Re}(\delta G(\eta,\sigma))|}{\int_{-R}^R d\sigma\ \sigma^{\ell_e-2} \mbox{Re}(\delta G(\eta,\sigma))}\le R^{\ell_o-\ell_e}\ll 1\ 
\ee
implying
\be\label{odd1}
\t_{\ell_o}> \t_{\ell_e}\ ,  \qquad \text{for all} \qquad \ell_o>\ell_e\ge2\ .
\ee
On the other hand, twists of two odd spin minimal twist operators do not obey any such condition. 

Odd spin minimal twist operators also satisfy a local convexity condition. For any odd $\ell_0\ge3$, the Cauchy-Schwarz inequality can be utilized to write  
\be\label{proof3}
0< \frac{\(\int_{-R}^R d\sigma\ \sigma^{\ell_o-2} \mbox{Re}(\delta G(\eta,\sigma))\)^2}{\int_{-R}^R d\sigma\ \sigma^{\ell_o-3} \mbox{Re}(\delta G(\eta,\sigma)) \int_{-R}^R d\sigma\ \sigma^{\ell_o-1} \mbox{Re}(\delta G(\eta,\sigma)) }\le 1\ .
\ee
In the regime $0<\eta\ll R\ll 1$,  we can use the sum rule (\ref{sum}) and the condition (\ref{odd1}) to obtain a local convexity condition for any odd $\ell_o\ge 3$
\be\label{odd11}
\t_{\ell_o}\ge \frac{\t_{\ell_o-1}+\t_{\ell_o+1}}{2}\ .
\ee
Note that this condition is stronger than (\ref{odd1}). Finally, we can apply this local convexity condition iteratively to obtain
\be\label{odd3}
\frac{\t_{\ell_o}-\t_{\ell_e}}{\ell_o-\ell_e}\ge \frac{\t_{\ell_e'}-\t_{\ell_e}}{\ell_e'-\ell_e}\ , \qquad 2\le \ell_e<\ell_o<\ell_e'
\ee
for all even $\ell_e$, $\ell_e'$ and odd $\ell_o$. In contrast to (\ref{convex}), the condition (\ref{odd3}) only provides a lower bound on $\t_{\ell_o}$ for $\ell_o\ge 3$ implying that the family of minimal twist operators with odd spins do not, in general, obey any global convexity condition. 

It is possible that there are a stronger bounds on odd spin minimal twist operators. For example, there may be  some relation between the continuous even spin and odd spin leading Regge trajectories above $\ell> 1$. We believe that such a relation, if it exists,  can be derived by combining our method with the Lorentzian inversion formula.

To summarize, we rederived and extended the convexity theorem of \cite{Komargodski:2012ek} by using analyticity and positivity properties of CFT correlators in Lorentzian signature. We end this section by commenting on a possible limitation of our argument. If the spectrum of operators that appears both in the OPE $\O\O^\dagger$ and $X\overline{X}$ have an accumulation point in the twist at $\tau_*$, it is not clear if our derivation is valid beyond $\t_{\ell}> \tau_*$.

\section{OPE of Nonidentical Operators}\label{section_mixed}
It is only natural to wonder if the preceding analysis can  be extended to OPEs of nonidentical scalar primaries in unitary CFTs. As the discussion  of section \ref{section_correlators} leads us to expect, the family of minimal twist operators appearing  in the OPE of nonidentical scalar primaries does obey some general constraints which we will discuss next. Before we proceed, we  introduce the notation
\begin{align}\label{define}
\delta \langle \O_I(\mathbf{1})\O_J(\rr)\O_K(-\rr)\O_L(-\mathbf{1})\rangle=& \langle   \O_I(\mathbf{1})\O_J(\rr)\O_K(-\mathbf{1})\O_L(-\rr)\rangle \nonumber\\
&-\langle \O_I(\mathbf{1})\O_J(\rr)\O_K(-\rr)\O_L(-\mathbf{1})\rangle \ ,
\end{align}
where operators are ordered as written. As remarked before, we do not need to cross any branch cuts to obtain the Lorentzian correlator $ \langle   \O_I(\mathbf{1})\O_J(\rr)\O_K(-\mathbf{1})\O_L(-\rr)\rangle$ and hence this correlators can be determined by using the Euclidean OPE even for $\r>1$. Clearly, the same argument holds for the Lorentzian correlator $ \langle   \O_J(\rr)\O_I(\mathbf{1})\O_L(-\rr)\O_K(-\mathbf{1})\rangle$ as well which implies that $ \langle   \O_I(\mathbf{1})\O_J(\rr)\O_K(-\mathbf{1})\O_L(-\rr)=\langle   \O_J(\rr)\O_I(\mathbf{1})\O_L(-\rr)\O_K(-\mathbf{1})\rangle\rangle$ even when  $\r>1$.

\subsection{Mixed Correlators and Rindler Positivity}

A Nachtmann-like theorem for nonidentical scalar primaries requires a condition similar to (\ref{con3}) for the mixed correlator $ \langle   \O_2^\dagger(\mathbf{1})\O_1(\rr)\O_2(-\rr)\O_1^\dagger(-\mathbf{1})\rangle$. In unitary Lorentzian QFTs, such a condition can be derived by utilizing Rindler positivity. Consider two operators $A$ and $B$ with support in the left Rindler wedge. We can define a positive inner product $(A,B)\equiv \langle B\overline{A}\rangle + \langle A\overline{B}\rangle$ by using Rindler positivity. This comes with the Cauchy-Schwarz identity
\be\label{CS}
(A,B)^2 \le (A,A)(B,B)\ .
\ee 

\subsubsection*{Inequality I}
Now consider two scalar primaries $\O_1$ and $\O_2$  with dimensions $\Delta_1$ and $\Delta_2$ respectively. First, we choose 
\begin{align}\label{snow1}
A&=(\r\br)^{\frac{\Delta_2+\delta_2}{2}}\O_1(\mathbf{1}) \O_2^\dagger(\rr)\pm (\r\br)^{\frac{\Delta_1+\delta_1}{2}}\O_2^\dagger(\mathbf{1})\O_1(\rr)\ , \nonumber\\
B&=\pm (\r\br)^{\frac{\Delta_1+\delta_1}{2}} \O_1(\rr)  \O_2^\dagger(\mathbf{1})+ (\r\br)^{\frac{\Delta_2+\delta_2}{2}}\O_2^\dagger(\rr)\O_1(\mathbf{1})\ ,
\end{align}
where $\delta_1$ and $\delta_2$ are arbitrary real numbers that we will fix later. The Cauchy-Schwarz identity (\ref{CS}) now leads to an inequality which applies to any unitary QFT. This inequality can be further simplified for CFTs where some of the correlators are related because of conformal symmetry. Specifically for $0<\br<1,\ \r>1$ we obtain\footnote{Note that we are using the notation (\ref{define}).}
\begin{align}\label{ineq1}
\pm 2 \mbox{Re}\ \delta \langle   \O_2^\dagger(\mathbf{1})\O_1(\rr)\O_2(-\rr)\O_1^\dagger(-\mathbf{1})\rangle \le (\r\br)^{\frac{\Delta_{21}+\delta_{21}}{2}} &\mbox{Re}\ \delta \langle  \O_1^\dagger(\mathbf{1}) \O_2(\rr)\O_2^\dagger(-\rr)\O_1(-\mathbf{1})\rangle\nonumber\\
+ (\r\br)^{-\frac{\Delta_{21}+\delta_{21}}{2}} \mbox{Re}\ \delta &\langle  \O_2^\dagger(\mathbf{1}) \O_1(\rr)\O_1^\dagger(-\rr)\O_2(-\mathbf{1})\rangle\
\end{align}
for any real $\delta_1$ and $\delta_2$, where  operators inside correlators are ordered as written. In the above equation, we have defined $\Delta_{21}=\Delta_2-\Delta_1$ and $\delta_{21}=\delta_2-\delta_1$. 

\subsubsection*{Inequality II}
Likewise, we can derive a similar inequality for a different mixed correlator by choosing
\begin{align}\label{snow2}
A&=(\r\br)^{\frac{\Delta_1+\delta_1}{2}}\O_1(\mathbf{1}) \O_1^\dagger(\rr)\pm (\r\br)^{\frac{\Delta_2+\delta_2}{2}}\O_2^\dagger(\mathbf{1})\O_2(\rr)\ , \nonumber\\
B&=(\r\br)^{\frac{\Delta_1+\delta_1}{2}} \O_1^\dagger(\rr)  \O_1(\mathbf{1}) \pm (\r\br)^{\frac{\Delta_2+\delta_2}{2}}\O_2(\rr)\O_2^\dagger(\mathbf{1})\ .
\end{align}
For $0<\br<1,\ \r>1$, the Cauchy-Schwarz identity (\ref{CS}) imposes
\begin{align}\label{ineq2}
\pm2 \mbox{Re}\ \delta \langle \O_2^\dagger(\mathbf{1})\O_2(\rr)  \O_1(-\rr)& \O_1^\dagger(-\mathbf{1})\rangle  \le  (\r\br)^{-\frac{\Delta_{21}+\delta_{21}}{2}} \mbox{Re}\ \delta \langle \O_1^\dagger(\mathbf{1}) \O_1(\rr) \O_1^\dagger(-\rr)  \O_1(-\mathbf{1}) \rangle \nonumber\\
&+ (\r\br)^{\frac{\Delta_{21}+\delta_{21}}{2}}\mbox{Re}\ \delta \langle \O_2^\dagger(\mathbf{1}) \O_2(\rr) \O_2^\dagger(-\rr)  \O_2(-\mathbf{1}) \rangle\
\end{align}
for any real $\delta_1$ and $\delta_2$. Of course, the above argument holds for any unitary QFT. However, conformal symmetry was used to simplify the Cauchy-Schwarz identity to obtain the final inequalities (\ref{ineq1}) and (\ref{ineq2}). So, for non-conformal QFTs, one must use the exact expression obtained from the   Cauchy-Schwarz identity (\ref{CS}) for $A$ and $B$ given in (\ref{snow1}) or (\ref{snow2}).

The inequalities (\ref{ineq1}) and (\ref{ineq2}) together, as we explain below, impose constraints on the family of minimal twist operators appearing in the OPE $\O_1\O_2$. To derive any such constraint, first  it is necessary to understand the Lorentzian lightcone limit of the mixed correlator $ \langle   \O_2^\dagger(\mathbf{1})\O_1(\rr)\O_2(-\rr)\O_1^\dagger(-\mathbf{1})\rangle$.

\subsection{Lightcone OPE}
It is straightforward to generalize the lightcone OPE of \cite{Hartman:2016lgu} for nonidentical scalar primaries. The contribution of the primary operator $\O^p_{\mu_1\mu_2\cdots}$ with conformal dimension $\Delta_p$ and spin $\ell_p$ and its descendants to the OPE  $\O_1\O_2$ in the lightcone limit $\r\br \rightarrow 0$ can be written as
\begin{align}\label{lope2}
\O_1(\rr)\O_2(-\rr)|_{\O_p} =\tilde{\lambda}_p \frac{\( \r \br\)^{\frac{\tau_p}{2} }\r^{\ell_p-1}}{(2\r\br)^{\frac{\Delta_{1}+\Delta_{2}}{2}}}  \int_{-\r}^\r d\r' \left(1-\frac{{\r'}}{\r} \right)^{\frac{-\Delta_{12}+h_p-2}{2}}\left(1+\frac{{\r'}}{\r} \right)^{\frac{\Delta_{12}+h_p-2}{2}} \O^p_{-\cdots -}(\r') 
\end{align}
where, $h_p=\Delta_p+\ell_p$ and twist $\tau_p=\Delta_p-\ell_p$. Actual value of the coefficient $\tilde{\lambda}_p$ is not important for our purpose, however let us note it here anyway
\be
\tilde{\lambda}_p=(-2)^{-\ell_p} \frac{ 2  c_{\O_1\O_2\O_p}  \Gamma \left(h_p\right)}{c_p \Gamma \left(\frac{1}{2} \left(h_p-\Delta _{21}\right)\right) \Gamma \left(\frac{1}{2} \left(h_p+\Delta _{21}\right)\right)}
\ee
 where, $c_p$ is the coefficient of the two-point function of $\langle \O_p\O_p\rangle$.\footnote{See \cite{Costa:2011mg} for conventions.} The lightcone OPE (\ref{lope2}) allows us to derive the contribution of the $\O_1\O_2\rightarrow\O_p\rightarrow \O_1^\dagger\O_2^\dagger$ conformal block to the correlator $\langle   \O_2^\dagger(\mathbf{1})\O_1(\rr)\O_2(-\rr)\O_1^\dagger(-\mathbf{1})\rangle$ (or any other related Lorentzian correlators) in the lightcone limit. In particular, in the Lorentzian lightcone limit $0<\eta\ll |\sigma|\ll 1$ we obtain\footnote{Variables $\eta$ and $\sigma$ are defined in (\ref{sigma}).}
\begin{align}\label{block2}
\delta \langle   \O_2^\dagger(\mathbf{1})\O_1(\rr)\O_2(-\rr)\O_1^\dagger(-\mathbf{1})\rangle|_{\O_p}=-i\frac{\eta^{\frac{1}{2}\(\tau_p-\Delta_{1}-\Delta_{2}\)}}{\sigma^{\ell_p-1}}\left(\tilde{C}_{\ell_p,\ell_p} +\sum_{n=1}^{\ell_p-1}\tilde{C}_{\ell_p,\ell_p-n}\sigma^n\right) 
\end{align}
plus correction terms that are either suppressed by higher powers of $\eta$ or decay for small $\sigma$. For our argument, as explained in appendix \ref{appendixB}, we can safely ignore these correction terms. In the above equation, actual values of coefficients $\tilde{C}_{\ell_p,\ell'}\propto |c_{\O_1\O_2\O_p}|^2$ are not important. These coefficients can be computed either from the OPE (\ref{lope2}) or by using the lightcone conformal block derived by Dolan and Osborn in \cite{Dolan:2000ut}. Finally, the mixed correlator in the Lorentzian lightcone limit  can  also be expressed as an asymptotic series which is organized by twist
\be
\delta \langle   \O_2^\dagger(\mathbf{1})\O_1(\rr)\O_2(-\rr)\O_1^\dagger(-\mathbf{1})\rangle \sim -i\sum_{\tau_p\le \tau_{cutoff}} \frac{\eta^{\frac{1}{2}\(\tau_p-\Delta_{1}-\Delta_{2}\)}}{\sigma^{\ell_p-1}}\left(\tilde{C}_{\ell_p,\ell_p} +\sum_{n=1}^{\ell_p-1}\tilde{C}_{\ell_p,\ell_p-n}\sigma^n\right)\ .
\ee

\subsection{Constraints on the Family of Minimal Twist Operators}
\begin{figure}
\begin{center}
\begin{tikzpicture}[baseline=-3pt, scale=1.80]
\begin{scope}[very thick,shift={(4,0)}]
\coordinate (v1) at (-1.5,-1.5) {};
\coordinate(v2) at (1.5,1.5) {};
\coordinate (v3) at (1.5,-1.5) {};
\coordinate(v4) at (-1.5,1.5) {};

\draw[thin,-latex]  (v1) -- (v2)node[left]{$x^+$};
\draw[thin,-latex]  (v3) -- (v4)node[right]{$\ x^-$};
\draw(-1.5,0)node[left]{ $\ \O_2^\dagger(\mathbf{1})$};
\draw(1.5,0)node[right]{ $ \O_1^\dagger(-\mathbf{1})$};
\filldraw[black]  (-1.5,0) circle (1pt);
\filldraw[black]  (1.5,0) circle (1pt);
\coordinate(v5) at (0,0) {};
\def \fac {.6};
\filldraw[black]  (-0.9,0.7) circle (1 pt);
\filldraw[black]  (0.9,-0.7) circle (1pt);
\draw(-0.9,0.7)node[left]{ $ \O_1(\rr)$};
\draw(0.9,-0.7)node[right]{ $ \O_2(-\rr)$};

\end{scope}
\end{tikzpicture}
\end{center}
\caption{\label{config2} \small A non-Rindler symmetric four-point function of two scalar primaries $\O_1$ and $\O_2$.}
\end{figure} 

We are now in a position to derive constraints on the family of minimal twist operators that appears in the OPE  $\O_1\O_2$. Consider the mixed Lorentzian correlator
\begin{align}
G_{mixed}(\eta,\sigma)=\eta^{\frac{1}{2}\(\Delta_1+\Delta_2+\delta_1+\delta_2\)} \langle   \O_2^\dagger(\mathbf{1})\O_1(\rr)\O_2(-\rr)\O_1^\dagger(-\mathbf{1})\rangle
\end{align}
in a unitary CFT in $d\ge3$ spacetime dimensions (see figure \ref{config2}). Moreover, in order to lighten the notation, we also define 
\begin{align}
 G_{IJ}(\eta,\sigma)=\eta^{\Delta_J+\delta_J} \langle  \O_I^\dagger(\mathbf{1}) \O_J(\rr)\O_J^\dagger(-\rr)\O_I(-\mathbf{1})\rangle\
\end{align}
with $I,J=1,2$. Similarly, we can define $\delta G_{mixed}(\eta,\sigma)$ and $\delta  G_{IJ}(\eta,\sigma)$ following (\ref{define}).  The positivity condition (\ref{con3}) again implies that $\mbox{Re}\ \delta G_{IJ}(\eta,\sigma)\ge 0$ for real $\sigma$ with $|\sigma|\ll 1$. Likewise,  inequalities (\ref{ineq1}) and (\ref{ineq2}) impose an upper bound on the real part of $\delta G_{mixed}(\eta,\sigma)$. In particular, the inequality (\ref{ineq1}) implies that  $2 |\mbox{Re}\ \delta G_{mixed}|\le \mbox{Re}\ (\delta G_{12}+\delta G_{21})$ for real positive $\sigma$. On the other hand, for negative $\sigma$ a similar bound can be obtained by using the covariant $i\epsilon$ prescription to relate $\mbox{Re} \delta G_{mixed}$ with the left hand side of (\ref{ineq2}). To summarize, correlators $G_{mixed}(\eta,\sigma)$ and $G_{IJ}(\eta,\sigma)$ for $0<\eta<1$ obey the following important properties:
\begin{enumerate}[label=(\alph*)]
\item \label{analytic2} As a function of complex $\sigma$, $G_{mixed}(\eta,\sigma)$ and $G_{IJ}(\eta,\sigma)$ are analytic near $\sigma \sim 0$ on the lower-half $\sigma$-plane. 
This follows from  the analyticity condition (\ref{con1}).

\item\label{bound3} The real part of the correlator $\delta G_{IJ}(\eta,\sigma)$ for real $\sigma$ with $|\sigma| < 1$ is non-negative.
 
\item \label{bound2} The real part of the correlator $\delta G_{mixed}(\eta,\sigma)$ for real $\sigma$ with $|\sigma| < 1$ is bounded 
\begin{align}\label{mixed}
&|\mbox{Re}\ \delta G_{mixed}(\eta,|\sigma|)| \leq \frac{1}{2}\mbox{Re}\(\delta G_{12}(\eta,|\sigma|)+\delta G_{21}(\eta,|\sigma|)\) \ , \nonumber\\
&|\mbox{Re}\ \delta G_{mixed}(\eta,-|\sigma|)| \leq  \frac{1}{2}\mbox{Re}\(\delta G_{11}(\eta,|\sigma|)+\delta G_{22}(\eta,|\sigma|)\)
\end{align}
for real but arbitrary $\delta_1$ and $\delta_2$, where the corrections on the right-hand sides  are suppressed by positive powers of both $\eta$ and $\sigma$.
\end{enumerate}

Conditions \ref{analytic2} and \ref{bound2} together suggest that in interacting CFTs in $d\ge 3$ dimensions there must be some relation between three families of minimal twist operators which we define next. First, consider all spin $\ell$ operators that appear in the OPE $\O_1\O_2$. Among these operators, we pick the lowest twist operator $\tilde{\O}^{(12)}_\ell$ which has twist $\t^{(12)}_\ell$. The set of operators $\{\tilde{\O}^{(12)}_\ell\ |\ \ell\in {\mathbb Z}^{\ge}\}$ is defined as the the family of minimal twist operators in the OPE $\O_1\O_2$. Similarly, we can define families of minimal twist operators for the OPE $\O_1 \O_1^\dagger$ and the OPE $\O_2 \O_2^\dagger$. Twists of these operators are denoted by  $\t^{(11)}_\ell$ and $\t^{(22)}_\ell$ respectively. Of course, there can be a distinct family of minimal twist operators with twist $\t^{(11\cap 22)}_\ell$ which appears both in $\O_1 \O_1^\dagger$ and $\O_2 \O_2^\dagger$, however, they always obey $\t^{(11\cap 22)}_\ell \ge \t^{(11)}_\ell,\t^{(22)}_\ell$.

\subsubsection*{Lower bound on $\t^{(12)}_\ell$ for even $\ell\ge 2$}
We can now use positivity properties \ref{bound3} and \ref{bound2} of CFT correlators to write the following relation for any two even $\ell_2\ge \ell_1\ge 2$
\be\label{diff1}
\frac{\int_{-R}^R d\sigma\ \sigma^{\ell_2-2} \mbox{Re}\delta G_{mixed}(\eta,\sigma)}{\int_{-R}^R d\sigma\ \sigma^{\ell_1-2} \sum_{I,J=1,2}\mbox{Re} \delta G_{IJ}(\eta,\sigma)}\le\frac{\int_{-R}^R d\sigma\ \sigma^{\ell_2-2} |\mbox{Re}\delta G_{mixed}(\eta,\sigma)|}{\int_{0}^R d\sigma\ \sigma^{\ell_1-2} \sum_{I,J=1,2}\mbox{Re} \delta G_{IJ}(\eta,\sigma)}\le \frac{1}{2}R^{\ell_2-\ell_1}< 1
\ee
implying that the quantity on the left hand side must not grow in the limit $\eta\rightarrow 0$.

As discussed in detail in appendices \ref{appendixA} and \ref{appendixB}, we can again define functions $\delta G_{mixed}^{(- )}(\eta,\sigma)$ and $\delta G_{IJ}^{(- )}(\eta,\sigma)$ which are analytic on the lower half $\sigma$-plane and have the property
\be
\mbox{Re}\ \delta G_{mixed}^{(- )}(\eta,\sigma)=\mbox{Re}\  \delta G_{mixed}(\eta,\sigma)\ , \qquad \mbox{Re}\ \delta G_{IJ}^{(- )}(\eta,\sigma)=\mbox{Re}\  \delta G_{IJ}(\eta,\sigma)
\ee
on the real line. The analyticity condition \ref{analytic2} now allows us to relate the line integrals of $\mbox{Re}\ \delta G_{mixed}(\eta,\sigma)$ and $\mbox{Re}\ \delta G_{IJ}(\eta,\sigma)$ to integrals over a semicircle by using the contour \ref{contour}. In the regime $0<\eta\ll R\ll 1$, we can use lightcone conformal blocks (\ref{block}) and (\ref{block2}) to calculate the correlators  $\delta G_{mixed}(\eta,\sigma)$ and $\delta G_{IJ}(\eta,\sigma)$ on the semicircle. The generalized correlators $\delta G_{mixed}^{(- )}(\eta,\sigma)$ and $\delta G_{IJ}^{(- )}(\eta,\sigma)$ are then obtained by simply removing all terms from $\delta G_{mixed}(\eta,\sigma)$ and $\delta G_{IJ}(\eta,\sigma)$ that decay for small $\sigma$ (see appendices \ref{appendixA} and \ref{appendixB}). Hence, the semicircle integrals can be further simplified by utilizing the identity (\ref{sachin}). In particular,  the leading contribution in the limit $0<\eta\ll R\ll 1$ can be obtained from  
\be\label{diff2}
\frac{\int_{-R}^R d\sigma\ \sigma^{\ell_2-2} \mbox{Re}\delta G_{mixed}(\eta,\sigma)}{\int_{-R}^R d\sigma\ \sigma^{\ell_1-2} \sum_{I,J=1,2}\mbox{Re} \delta G_{IJ}(\eta,\sigma)}=\frac{\sum_{\ell'\ge \ell_2}\tilde{C}_{\ell', \ell_2}\ \eta^{\frac{1}{2}\(\tilde{\tau}^{(12)}_{\ell'}+\delta_{1}+\delta_{2}\)}}{C^{(11)}_{\ell_1, \ell_1}\eta^{\frac{1}{2}\tilde{\tau}^{(11)}_{\ell_1}+\delta_{1}}+C^{(22)}_{\ell_1, \ell_1}\eta^{\frac{1}{2}\tilde{\tau}^{(22)}_{\ell_1}+\delta_{2}}}
\ee
where we have ignored terms with $\eta^{\frac{1}{2}\t^{(11\cap 22)}_\ell}$  which never dominate because  $\t^{(11\cap 22)}_\ell \ge \t^{(11)}_\ell,\t^{(22)}_\ell$. In the above equation, coefficients $\tilde{C}_{\ell', \ell_2}$, $C^{(11)}_{\ell_1, \ell_1}$, and $C^{(22)}_{\ell_1, \ell_1}$ do not depend on $\eta$ and hence this equation can be consistent with (\ref{diff1}) for any two even $\ell_2\ge \ell_1\ge 2$ if and only if $\tilde{\tau}^{(12)}_{\ell_1} \ge \mbox{Min}[\tilde{\tau}^{(11)}_{\ell_1}+\delta_1-\delta_2, \tilde{\tau}^{(22)}_{\ell_1}+\delta_2-\delta_1]$ for any real $\delta_1$ and $\delta_2$. The optimal bound is obtained for $\delta_1-\delta_2=(\tilde{\tau}^{(22)}_{\ell_1}-\tilde{\tau}^{(11)}_{\ell_1})/2$ implying
\be\label{even2}
\tilde{\tau}^{(12)}_{\ell}\ge \frac{1}{2}\(\tilde{\tau}^{(11)}_{\ell}+\tilde{\tau}^{(22)}_{\ell} \)
\ee
for any even $\ell \ge 2$.

\subsubsection*{Lower bound on $\t^{(12)}_\ell$ for odd $\ell\ge 3$}
Clearly, the preceding argument applies even when $\ell_2$ is odd implying $\tilde{\tau}^{(12)}_{\ell_o}\ge \frac{1}{2}(\tilde{\tau}^{(11)}_{\ell_o-1}+\tilde{\tau}^{(22)}_{\ell_o-1} )$ for any odd $\ell_o\ge 3$. Furthermore, this condition enables us to derive a stronger lower bound. For any odd $\ell_o\ge3$, the Cauchy-Schwarz inequality of integrable functions allows us to write\footnote{We are using the following notation
\be
\delta G^{\delta_1,\delta_2}_{mixed}(\eta,\sigma)\equiv \eta^{\frac{1}{2}\(\Delta_1+\Delta_2+\delta_1+\delta_2\)} \delta \langle   \O_2^\dagger(\mathbf{1})\O_1(\rr)\O_2(-\rr)\O_1^\dagger(-\mathbf{1})\rangle\ . \nonumber
\ee}  
\begin{align}\label{diff2}
&\(\int_{-R}^R d\sigma\ \sigma^{\ell_o-2} \mbox{Re}\ \delta G^{\delta_1+\delta_1',\delta_2+\delta_2'}_{mixed}(\eta,\sigma)\)^2\nonumber\\
&\qquad \le \(\int_{-R}^R d\sigma\ \sigma^{\ell_o-3} \mbox{Re}\ \delta G^{2\delta_1,2\delta_2}_{mixed}(\eta,\sigma)\)\( \int_{-R}^R d\sigma\ \sigma^{\ell_o-1}\mbox{Re}\ \delta G^{2\delta_1',2\delta_2'}_{mixed}(\eta,\sigma) \)
\end{align}
for any real $\delta_1,\delta_2,\delta_1',\delta_2'$. The rest of the argument is exactly the same as before. In the regime $0<\eta\ll R\ll 1$,  inequalities (\ref{diff1}) and (\ref{diff2}) impose a lower bound on $\tilde{\tau}^{(12)}_{\ell_o}$ that depends on $\delta_1,\delta_2,\delta_1',\delta_2'$. By choosing $\delta_1-\delta_2=(\tilde{\tau}^{(22)}_{\ell_o-1}-\tilde{\tau}^{(11)}_{\ell_o-1})/4$ and $\delta_1'-\delta_2'=(\tilde{\tau}^{(22)}_{\ell_o+1}-\tilde{\tau}^{(11)}_{\ell_o+1})/4$ we obtain a  local convexity condition for $\tilde{\tau}^{(12)}_{\ell_o}$ 
\be\label{odd2}
\tilde{\tau}^{(12)}_{\ell_o}\ge \frac{1}{2}\(\frac{\tilde{\tau}^{(11)}_{\ell_o-1}+\tilde{\tau}^{(22)}_{\ell_o-1}}{2}+\frac{\tilde{\tau}^{(11)}_{\ell_o+1}+\tilde{\tau}^{(22)}_{\ell_o+1} }{2}\)\ ,
\ee
for any odd $\ell_o\ge 3$. Similar to the preceding section, formally bounds (\ref{even2}) and (\ref{odd2}) are valid up to the first  twist accumulation point of the $\O_1\O_2$ OPE. 

The above constraints suggest that there are some general relations among various continuous spin leading Regge trajectories.  It is possible that such relations can be derived by generalizing the analysis of \cite{Costa:2017twz} for mixed correlators.

\section{3d Ising Model and Other Examples}\label{section_ising}
In this section we provide some simple examples to demonstrate that unitary CFTs in $d\ge 3$ dimensions indeed obey the above constraints. 
\subsection{3d Ising CFT}
The first example that we consider is the 3d Ising CFT. This CFT contains operators $\sigma$ and $\epsilon$ which are the lowest-dimension $\mathbb{Z}_2$-odd and $\mathbb{Z}_2$-even scalar primaries of the theory, respectively. In recent years, numerical bootstrap methods have led to significant progress in constraining the  data of the 3d Ising CFT. For instance, the bootstrap has provided precise  conformal dimensions for operators $\sigma$ and $\epsilon$ just from crossing symmetry and unitarity \cite{ElShowk:2012ht,El-Showk:2014dwa,Kos:2014bka,Simmons-Duffin:2015qma,Kos:2016ysd,Simmons-Duffin:2016wlq}. Furthermore, the same principles, as demonstrated in \cite{Simmons-Duffin:2016wlq}, are also sufficient to determine the spectrum of the 3d Ising CFT numerically in a systematic way. In fact, the numerical data of \cite{Simmons-Duffin:2016wlq} is so precise for several low-lying operators that we can actually study and compare families of minimal twist operators that appear in the OPE of $\sigma\sigma$, $\epsilon\epsilon$, and $\sigma\epsilon$.

\begin{figure}
\hfill
\subfigure[]{\includegraphics[width=7.5cm]{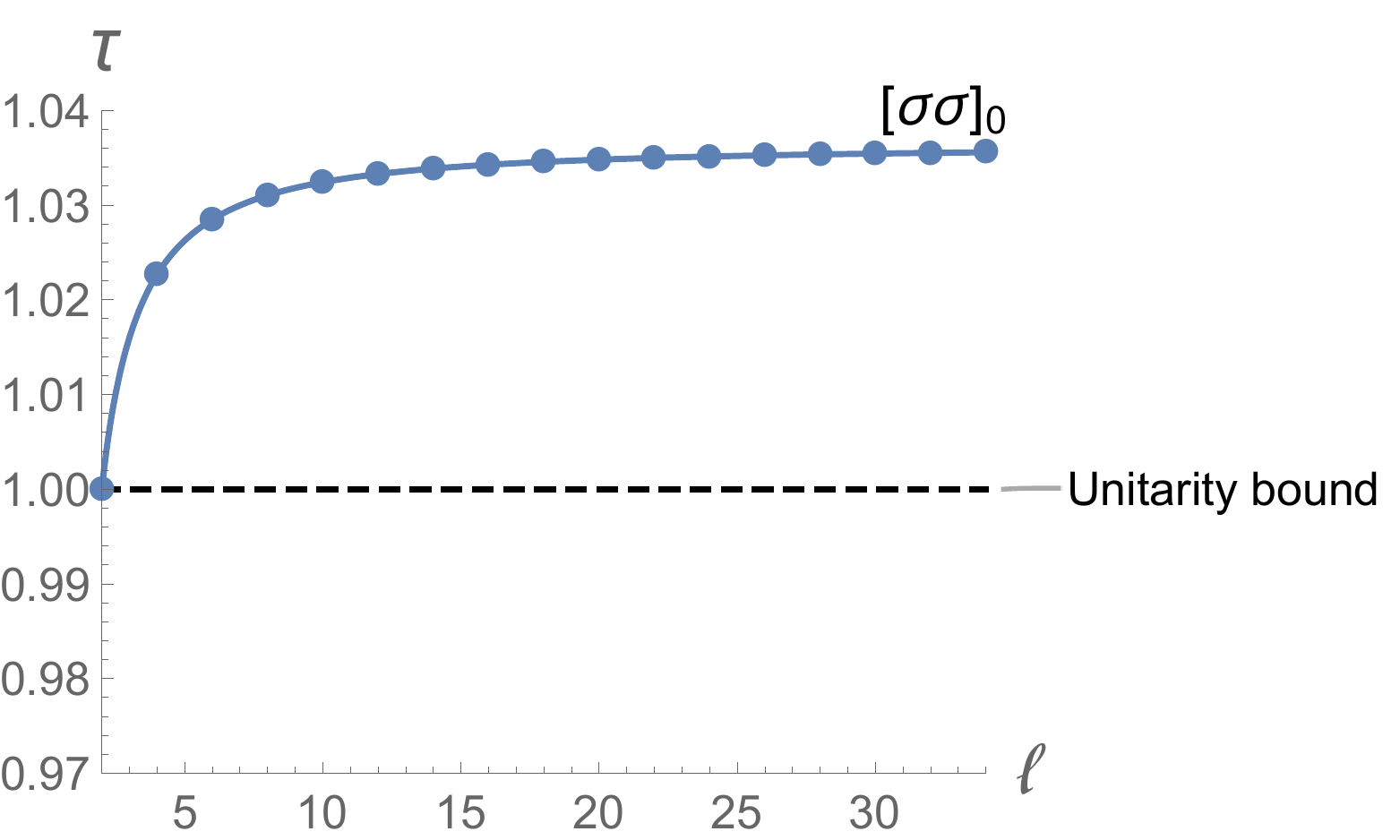}}
\hfill
\subfigure[]{\includegraphics[width=8cm]{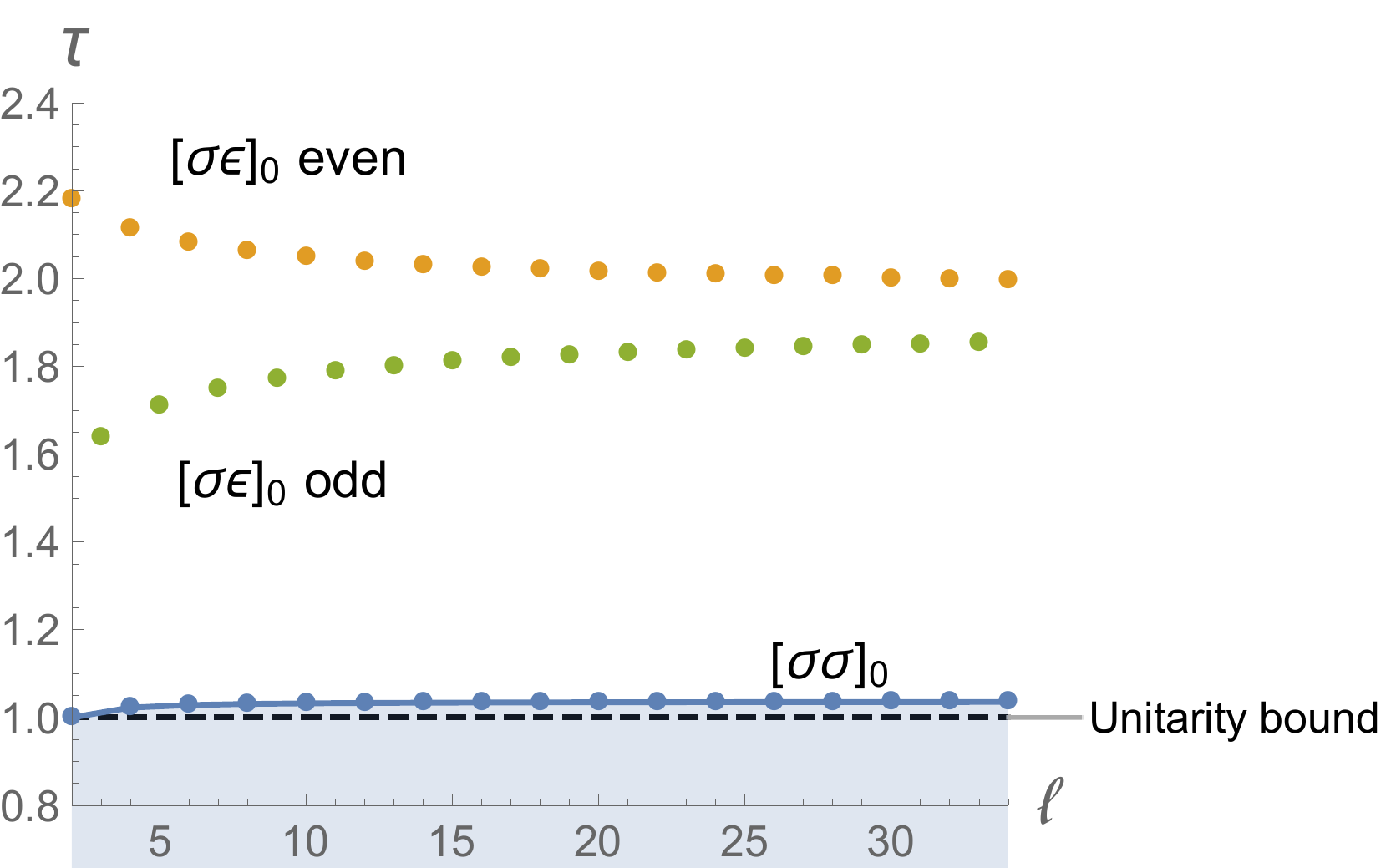}}
\hfill
\caption{\small Families of minimal twist operators in 3d Ising CFT. In figure (a), using the numerical data of \cite{Simmons-Duffin:2016wlq} we show that the family $[\sigma\sigma]_0$ indeed obeys the Nachtmann theorem. Moreover, the bounds (\ref{even2}) and (\ref{odd2}) imply that the shaded region in figure (b) is ruled out for the $[\sigma\epsilon]_0$ family. The numerical results for even and odd spin operators in the $[\sigma\epsilon]_0$ family are shown in yellow and green respectively.}\label{ising}
\end{figure}

Let us now examine the numerical 3d Ising data of \cite{Simmons-Duffin:2016wlq}. The set of operators $[\sigma\sigma]_0$ is of particular importance since these operators form the family of minimal twist operators for both $\sigma\sigma$ and $\epsilon\epsilon$ OPEs. Clearly, the numerical data of \cite{Simmons-Duffin:2016wlq} for the family $[\sigma\sigma]_0$ is consistent with the Nachtmann theorem (see figure \ref{ising}). Moreover, as we explained in the last section, the $[\sigma\sigma]_0$ family also provides a lower bound for the twists of $[\sigma\epsilon]_0$ operators (both even and odd spins) which are minimal twist operators for the $\sigma \epsilon$ OPE. This lower bound is stronger than the unitarity bound, however, it is still relatively weak. Of course, the numerical data of \cite{Simmons-Duffin:2016wlq} for the family $[\sigma\epsilon]_0$, as we show in figure \ref{ising}, is consistent with analytic bounds (\ref{even2}) and (\ref{odd2}).

Interestingly, both even and odd spin operators in the family $[\sigma\epsilon]_0$ exhibit some nice features. For example, the 3d Ising data suggests that the twists of odd (or even) spin operators  in the family $[\sigma\epsilon]_0$  obey some monotonicity and convexity conditions. However, we believe that these conditions are not true in generic CFTs.

\subsection{Large Spin Bootstrap}
\subsubsection*{Real Scalars}
Constraints (\ref{even2}) and (\ref{odd2}) are also visible from the large spin bootstrap of real scalar primaries \cite{Komargodski:2012ek,Fitzpatrick:2012yx}. Consider the CFT correlator $\langle \phi_1(x_1)\phi_1(x_2)\phi_2(x_3)\phi_2(x_4)\rangle$ of two real scalar primaries $\phi_1$ and $\phi_2$ with dimensions $\Delta_1$ and $\Delta_2$, respectively. The crossing relation in the traditional lightcone limit can be approximated as 
\be
\blockS{\phi_1}{\phi_1}{\phi_2}{\phi_2}{\mathbf 1} + \blockS{\phi_1}{\phi_1}{\phi_2}{\phi_2}{\O_m}
 \approx \sum_{n,\ell} \blockT{\phi_1}{\phi_1}{\phi_2}{\phi_2}{[\phi_1\phi_2]_{n,\ell}}  \ ,
\ee
where each diagram represents a conformal block and $\O_m$ is the lowest twist operator that appears both in $\phi_1\phi_1$ and $\phi_2\phi_2$ OPEs. Twists $\tau^{[\phi_1\phi_2]}_{n,\ell}$ of the double twist operators $[\phi_1\phi_2]_{n,\ell}$ for large spin can be obtained by solving the above crossing equation. In particular, for the minimal twist tower (both even and odd spins) at large spin we get \cite{Li:2015rfa}
\be
\tau^{[\phi_1\phi_2]}_{0,\ell}=\Delta_1+\Delta_2- \xi^{\O_m}_{\Delta_1,\Delta_2}\frac{c_{\phi_1\phi_1\O_m}c_{\phi_2\phi_2\O_m}}{2^{\ell_m}\ell^{\tau_m}}\ , \quad \xi^{\O_m}_{\Delta_1,\Delta_2}= \frac{2\Gamma\(\tau_m+2\ell_m\)}{\Gamma\(\frac{\tau_m}{2}+\ell_m\)^2}\prod_{i=1,2} \frac{\Gamma\(\Delta_i\)}{\Gamma\(\Delta_i-\frac{\tau_m}{2}\)}
\ee
where $\tau_m$ is the twist and $\ell_m$ is the spin of $\O_m$ and $c_{\phi_1\phi_1\O_m}, c_{\phi_2\phi_2\O_m}$ are OPE coefficients. Similarly, by considering the u-channel, for large spin we obtain 
\be
\tau^{[\phi_1\phi_1]}_{0,\ell}=2\Delta_1- \xi^{\O_m}_{\Delta_1,\Delta_1}\frac{c_{\phi_1\phi_1\O_m}^2}{2^{\ell_m}\ell^{\tau_m}}\ , \qquad \tau^{[\phi_2\phi_2]}_{0,\ell}=2\Delta_2- \xi^{\O_m}_{\Delta_2,\Delta_2}\frac{c_{\phi_2\phi_2\O_m}^2}{2^{\ell_m}\ell^{\tau_m}}
\ee
implying $2\tau^{[\phi_1\phi_2]}_{0,\ell} \ge \tau^{[\phi_1\phi_1]}_{0,\ell}+\tau^{[\phi_2\phi_2]}_{0,\ell} $ which agrees with (\ref{even2}) and (\ref{odd2}) at large $\ell$. In fact, if we consider exchange of multiple operators in the t-channel, contributions of each such operator to $\tau^{[\phi_1\phi_2]}_{0,\ell}$, $\tau^{[\phi_1\phi_1]}_{0,\ell}$, and $\tau^{[\phi_2\phi_2]}_{0,\ell} $ satisfy the above inequality individually. 

\subsubsection*{Complex Scalars}
All the bounds discussed in this paper can be nicely demonstrated by studying the large spin behaviors of various double twist operators of a complex scalar primary $\O$ which is charged under a global $U(1)$ symmetry. This scenario  has been analyzed in detail in  \cite{Komargodski:2012ek, Li:2015rfa}. Consider the correlator $\langle \O(x_1)\O^\dagger(x_2)\O(x_3)\O^\dagger(x_4)\rangle$. The crossing equation in the lightcone limit has the general form 
\begin{align}
{\small \blockS{\O}{\O^\dagger}{\O^\dagger}{\O}{\mathbf 1} + \blockTT{\O}{\O^\dagger}{\O^\dagger}{\O}{S+J+T}
 \approx \sum_{n,\ell} \blockT{\O}{\O^\dagger}{\O^\dagger}{\O}{[\O\O^\dagger]_{n,\ell}}\ ,}
\end{align}
where $T$ is the stress tensor, $J$ is the $U(1)$ symmetry current and $S$ is a low dimensional scalar (if present). Similarly, we can write a slightly different crossing equation in the lightcone limit
\begin{align}
{\small \blockS{\O}{\O^\dagger}{\O}{\O^\dagger}{\mathbf 1} + \blockTT{\O}{\O^\dagger}{\O}{\O^\dagger}{S+J+T}
 \approx \sum_{n,\ell} \blockT{\O}{\O^\dagger}{\O}{\O^\dagger}{[\O\O]_{n,\ell}}\ ,}
\end{align}
which already suggests that twists of $[\O\O]_{n,\ell}$ and $[\O\O^\dagger]_{n,\ell}$ are not completely unrelated at large spin. Finally, let us also include the possibility of a low dimensional charged scalar $C$ that can appear in the $\O\O$ OPE. This leads to another crossing relation in the lightcone limit 
\begin{align}
{\small \blockS{\O}{\O}{\O^\dagger}{\O^\dagger}{C}
 \approx \sum_{n,\ell} \blockT{\O}{\O}{\O^\dagger}{\O^\dagger}{[\O\O^\dagger]_{n,\ell}}\ .}
\end{align}
One can solve these crossing equations simultaneously to obtain twists $\tau^{[\O\O^\dagger]}_{n,\ell}$ and $\tau^{[\O\O]}_{n,\ell}$ of the double twist operators $[\O\O^\dagger]_{n,\ell}$ and $[\O\O]_{n,\ell}$, respectively,  at large $\ell$. First, we focus on the minimal twist tower $[\O\O^\dagger]_{0,\ell}$. Because of the charged scalar $C$, even and odd spin operators in $[\O\O^\dagger]_{0,\ell}$ behave differently. In particular, in $d$ spacetime dimensions following \cite{Li:2015rfa}  one can obtain
\be\label{complex1}
\tau^{[\O\O^\dagger]}_{0,\ell^{\pm}}\approx2\Delta_\O-\frac{1}{\ell^{d-2} S_d^2}\(\frac{d^2 \Delta_\O^2 \xi^{T}_{\Delta_\O,\Delta_\O}}{4(d-1)^2 C_T }+\frac{\xi^{J}_{\Delta_\O,\Delta_\O}}{2C_J }\)-\frac{|c_{\O\O^\dagger S}|^2  \xi^{S}_{\Delta_\O,\Delta_\O}}{\ell^{\Delta_S}}\mp \frac{|c_{\O\O^\dagger C}|^2  \xi^{C}_{\Delta_\O,\Delta_\O}}{\ell^{\Delta_C}}
\ee
for large even ($\ell^+$) and odd ($\ell^-$) spins. In the above equation, we have used the notation of \cite{Li:2015rfa} where $C_J$ and $C_T$ are central charges and $S_d=\frac{2 \pi^{d/2}}{\Gamma(d/2)}$. Note that $\xi^{\O_m}_{\Delta_1,\Delta_2}$ is a positive quantity and hence $\tau^{[\O\O^\dagger]}_{0,\ell}$ for even spin is consistent with the Nachtmann theorem. Whereas, for odd spin $\tau^{[\O\O^\dagger]}_{0,\ell}$ does not in general have to be a monotonically increasing convex function of $\ell$. Moreover, the equation (\ref{complex1}) also implies that for large spin $\tau^{[\O\O^\dagger]}_{0,\ell^-}\ge \tau^{[\O\O^\dagger]}_{0,\ell^+}$ which is consistent with (\ref{odd11}).

Similarly, for the minimal twist tower $[\O\O]_{0,\ell}$ at large spin we obtain \cite{Li:2015rfa}
\be
\tau^{[\O\O]}_{0,\ell}\approx2\Delta_\O-\frac{1}{\ell^{d-2} S_d^2}\(\frac{d^2 \Delta_\O^2 \xi^{T}_{\Delta_\O,\Delta_\O}}{4(d-1)^2 C_T }-\frac{\xi^{J}_{\Delta_\O,\Delta_\O}}{2C_J }\)-\frac{|c_{\O\O^\dagger S}|^2  \xi^{S}_{\Delta_\O,\Delta_\O}}{\ell^{\Delta_S}}
\ee
implying that $\tau^{[\O\O]}_{0,\ell}$  does not obey a Nachtmann-like theorem in general. However, $\tau^{[\O\O]}_{0,\ell}$ is bounded from below by a monotonically increasing convex function of $\ell$: $\tau^{[\O\O]}_{0,\ell}\ge \tau^{[\O\O^\dagger]}_{0,\ell^+}$ which is consistent with (\ref{even2}). This is easy to understand in the context of the AdS/CFT correspondence where anomalous dimensions correspond to  binding energies between two well separated rotating charged particles. In the bulk, $U(1)$ gauge interactions result in a repulsive force between like-charged particles but an attractive force between  oppositely charged particles. This immediately implies that anomalous dimensions of $[\O\O^\dagger]_{0,\ell^+}$ cannot be larger than anomalous dimensions of $[\O\O]_{0,\ell}$.

The leading order lightcone bootstrap results can be extended  to all orders in inverse powers of the spin, and for all twists by using the large spin perturbation theory framework of \cite{Alday:2015eya,Alday:2015ota,Alday:2016njk}. Our non-perturbative results are completely consistent with perturbation theory results.

\section{Regge Limit and Large $N$ CFTs}\label{section_regge}
We now consider another intrinsically Lorentzian limit of a CFT four-point function -- the CFT Regge limit. Our starting point is again the  corrrelator $\langle O_4(\mathbf{1}) O_1(\rr) O_2(-\rr) O_3(-\mathbf{1})  \rangle$ where operators are ordered as written  with $1>\br>0$ and $\rho>1$. Similar to the Lorentzian lightcone limit, operator pairs $O_4(x_4),O_1(x_1)$ and $O_2(x_2),O_3(x_3)$ are time-like separated. This Lorentzian correlator is obtained from the Euclidean correlator by analytically continuing $\r$ along the path shown in figure \ref{dt1}. The CFT Regge limit is then defined by \cite{Brower:2006ea,Cornalba:2007fs,Cornalba:2008qf,Costa:2012cb}
\be\label{regge2}
\sigma\rightarrow 0\ , \qquad \text{with} \qquad \eta=\text{fixed}
\ee
of the Lorentzian correlator $\langle O_4(\mathbf{1}) O_1(\rr) O_2(-\rr) O_3(-\mathbf{1})  \rangle$, where $\sigma$ and $\eta$ are defined in (\ref{sigma}). Clearly, the Lorentzian lightcone limit is a special case of the Regge limit. 

The main point we wish to emphasize in this section is that equations (\ref{proof1}), (\ref{proof2}), (\ref{proof3}) as well as equations (\ref{diff1}), (\ref{diff2}) are valid even in the regime $0<R\ll \eta<1$. Just like before, the analyticity condition of CFT correlators in Lorentzian signature now constraints the Regge behavior of certain CFT correlators by relating $\sigma$-integrals on the real line to an integral of the Regge limit of CFT correlators over the semicircle. However, these constraints are expected to be theory dependent because the Regge limit, for finite $\eta$, is dominated by high spin exchanges which are non-universal. 

For the purpose of demonstration,  we circumvent the intricacies of  the Regge limit by assuming a specific Regge behavior. We consider CFTs in which the correlator $\langle O_1(\mathbf{1}) O_2(\rr)\overline{O_2(\rr)}\ \overline{O_1(\mathbf{1})} \rangle$ in the Regge limit admits an  expansion
\be\label{con_anc}
\frac{\langle O_1(\mathbf{1}) O_2(\rr)\overline{O_2(\rr)}\ \overline{O_1(\mathbf{1})} \rangle}{\langle O_1(\mathbf{1})  \overline{O_1(\mathbf{1})} \rangle\langle O_2(\rr)\overline{O_2(\rr)}\rangle} \sim 1+ i \sum_{L=1,2,\cdots}  \frac{c_L}{ \sigma^{L-1}}\ , \qquad \frac{1}{\Lambda}\ll |\sigma|\ll \eta<1
\ee
for some operators $O_1$ and $O_2$ with or without spin, where $\Lambda$ is some cut-off scale and $c_L$ are $\sigma$ independent real coefficients. This happens naturally in large-$N$ CFTs. At first sight, the expansion (\ref{con_anc}) for $L>2$ appears to be in contradiction with the chaos bound \cite{Maldacena:2015waa,Afkhami-Jeddi:2016ntf}. This suggests that coefficients $c_L$ are highly constrained. Alternatively, relations  (\ref{proof1}), (\ref{proof2}) and (\ref{proof3}) impose constraints on $c_L$. These constraints ensure that the expansion (\ref{con_anc}) is consistent with the chaos bound. 

Let us now be more precise. First, conditions \ref{analytic} and \ref{bound} lead to a positivity condition  
\be
c_L\ge 0\ ,  \qquad \text{for} \qquad \text{even}\ L\ge 2\ .
\ee
Moreover, the condition  \ref{bound}  along with the relation  (\ref{proof1})  in the limit $\frac{1}{\Lambda}\ll R\ll \eta<1$ also imply the parametric bound 
\be
\frac{|c_{L+1}| }{c_{L}} \lesssim \frac{1}{\Lambda} \ , \qquad \frac{c_{L+2}}{c_{L}} \lesssim \frac{1}{\Lambda^2}\ 
\ee
for any even $L\ge 2$. Similarly, equations (\ref{proof2}) and (\ref{proof3}) in the limit $\frac{1}{\Lambda}\ll R\ll \eta<1$ lead to the following quadratic relations
\be
\(c_{L+2}\)^2\le c_{L}c_{L+4}\ , \qquad \(c_{L+1}\)^2\le c_{L}c_{L+2} \qquad \text{for even} \quad L\ge 2\ .
\ee
Therefore, all coefficients with $L>2$ must be parametrically suppressed in a systematic way implying that terms in (\ref{con_anc}) that grow faster than $1/\sigma$ can never dominate for $1\gg |\sigma|\gg \frac{1}{\Lambda}$. This makes the expansion (\ref{con_anc}) consistent with the chaos bound. The above constraints are particularly useful in the context of the AdS/CFT correspondence where these constraints should be interpreted as bounds on various interactions of low energy effective field theories in AdS from UV consistency.

Of course, the discussion of this section can be extended to the mixed correlator $\langle   O_2^\dagger(\mathbf{1})O_1(\rr)O_2(-\rr)O_1^\dagger(-\mathbf{1})\rangle$ simply by following the discussion of section \ref{section_mixed}. If the mixed correlator admits an expansion similar to (\ref{con_anc}) in the Regge limit, one can derive analogous bounds by exploiting equations (\ref{diff1}) and (\ref{diff2})  in the regime $\frac{1}{\Lambda}<R\ll \eta<1$.

~\\

This concludes our discussion of various generalizations of the Nachtmann theorem in CFT.
%%%%%%%%%%%%%%%%%%%%%%%%%%%%%
\section*{Acknowledgments}
It is my pleasure to thank  Nima Afkhami-Jeddi, Luis  Alday, Tom Hartman, Jared Kaplan, David Meltzer, Joao Penedones,  Slava Rychkov,  Zahra Zahraee, and Alexander Zhiboedov for helpful discussions.  I was supported in part by the Simons Collaboration Grant on the Non-Perturbative Bootstrap. I am  grateful to Perimeter Institute for Theoretical Physics and the Simons Bootstrap Collaboration for hospitality and support during the Bootstrap 2019 workshop where part of this work was completed. I also thank  the Simons Center for Geometry and Physics of Stony Brook University for providing additional support during the workshop on Developments in the Numerical Bootstrap where some of this work was done.

%%%%%%%%%%%%%%%%%%%%%%%%%%%%%%%
\appendix

\section{A Detailed Derivation of the Sum Rule}\label{appendixA}

We now provide a complete derivation of the sum rule (\ref{sum}) by emphasizing some of the important points. The main argument is simple when all exchanged operators have integer dimensions. The argument is more subtle when operators with non-integer dimensions are exchanged.  The dots in  equation (\ref{block}) contain terms that decay with a non-integer but positive powers of $\sigma$ when operators with non-integer dimensions are exchanged. These terms lead to additional contributions which appear to spoil the sum rule (\ref{sum}). However, these contributions can be  subtracted by analytically continuing $G_0(\eta,\sigma)$ appropriately, as we explain next.

First, we use the OPE (\ref{lope}) to derive the contribution of the $\O\O^\dagger\rightarrow\O_p\rightarrow X\overline{X}$ conformal block to the correlator
\be
G_0(\eta, \sigma)=\frac{\langle X(\mathbf{1}) \O(\rr) \overline{X(\mathbf{1})}\ \overline{\O(\rr)} \rangle}{\langle X(\mathbf{1})\overline{X(\mathbf{1})} \rangle \langle \O(\rr)\overline{\O(\rr)}\rangle}
\ee
in the  lightcone limit. Clearly, the correlator $G_0(\eta, \sigma)$ is well defined only for real $\sigma$. In particular, for real positive $\sigma\ll 1$, we can write the following expansion in the lightcone limit 
\begin{align}\label{app1}
G_0(\eta,\sigma)|_{\O_p}=\eta^{\frac{\tau_p}{2}} \sigma^{\Delta_p}\(\sum_{n=0,2,4,\cdots} C^{(p)}_{n} \sigma^n\)+\cdots\ ,
\end{align}
where $C^{(p)}_{n}$ are real coefficients and dots represent terms that are suppressed by higher powers of $\eta$. Obviously, the correction terms are suppressed by positive powers of $\sigma$ as well. Moreover, notice that for negative $\sigma$
\be
G_0(\eta,\sigma)|_{\O_p}=(-1)^{\ell_p}G_0(\eta,|\sigma|)|_{\O_p}\ .
\ee
Next, we consider the correlator
\be
G(\eta, \sigma)=\frac{\langle X(\mathbf{1}) \O(\rr)\overline{\O(\rr)}\ \overline{X(\mathbf{1})} \rangle}{\langle X(\mathbf{1})\overline{X(\mathbf{1})} \rangle \langle \O(\rr)\overline{\O(\rr)}\rangle}
\ee 
in the lightcone limit. The goal is to figure out the behavior of $G(\eta,\sigma)$ for complex $\sigma$ with $|\sigma|<1$. The contribution of the $\O\O^\dagger\rightarrow\O_p\rightarrow X\overline{X}$ conformal block to the above correlator in the Lorentzian lightcone limit can be computed using the OPE (\ref{lope}). For real positive $\sigma\ll 1$, this contribution has the following structure 
\be\label{app2}
G(\eta,\sigma)|_{\O_p}=G_0(\eta,\sigma)|_{\O_p}+\(G^{(p)}_{int}(\eta,\sigma)+G^{(p)}_{nint}(\eta,\sigma)\)+\(\delta G^{(p)}_{int}(\eta,\sigma)+\delta G^{(p)}_{nint}(\eta,\sigma)\)
\ee
where the real part of $G(\eta,\sigma)|_{\O_p}$ is exactly $G_0(\eta,\sigma)|_{\O_p}$ which  is given in (\ref{app1}). The rest of the terms in the above equation, for real positive $\sigma\ll 1$, are completely imaginary. This follows from the fact that $G(\eta,\sigma)|_{\O_p}-G_0(\eta,\sigma)|_{\O_p}$ can be written as an integral of the discontinuity of some correlator across a branch cut in the $\r$-plane. The leading imaginary contribution in the lightcone limit has two distinct parts $G^{(p)}_{int}(\eta,\sigma)$ and $G^{(p)}_{nint}(\eta,\sigma)$. The contribution $G^{(p)}_{int}(\eta,\sigma)$ which only has integer powers of $\sigma$, grows for small $\sigma$s
\be\label{bck1}
G^{(p)}_{int}(\eta,\sigma)=-i\frac{\eta^{\frac{\tau_p}{2}}}{\sigma^{\ell_p-1}}\sum_{n=0,2,4,\cdots}C_{\ell_p,\ell_p-n}\sigma^n\ . 
\ee
This is the contribution that dominates in the Lorentzian lightcone limit, however, there can be other terms with non-integer powers\footnote{For simplicity, we are assuming that all exchanged operators have non-integer dimensions. Of course, for integer $\Delta_p$, one can take the integer limit at the end.} of $\sigma$ that decay for small $\sigma$
\begin{align}\label{block1}
G^{(p)}_{nint}(\eta,\sigma)=i \eta^{\frac{\tau_p}{2}} \sigma^{\Delta_p}\sum_{n=0,2,4,\cdots}\tilde{C}^{(p)}_{n} \sigma^n\ .
\end{align}
In fact, later we will argue that $G^{(p)}_{nint}(\eta,\sigma)$ must be present in order to make $G(\eta,\sigma)$ analytic on the lower half complex-$\sigma$ plane. Finally, $\delta G^{(p)}_{int}(\eta,\sigma)$ and $\delta G^{(p)}_{nint}(\eta,\sigma)$ represent terms that are suppressed by higher powers of $\eta$. These correction terms are more difficult to compute since they depend on higher order terms of the lightcone OPE (\ref{lope}). Nonetheless, it is easy to estimate the general behaviors of these correction terms. First of all, conformal invariance dictates that the correction terms with integer powers of $\sigma$ cannot grow faster than $1/\sigma^{\ell-1}$. On the other hand,  correction terms with non-integer powers of $\sigma$ are fixed by the correction terms in (\ref{app1}) from analyticity and crossing. Thus, we conclude that 
\be
\delta G^{(p)}_{int}(\eta,\sigma)=\O\(i \frac{\eta^{\frac{\tau_p}{2}+1}}{ \sigma^{\ell-1}}\)\ , \qquad \delta G^{(p)}_{int}(\eta,\sigma)=\O\(i \eta^{\frac{\tau_p}{2}+1}\sigma^a\)\ ,
\ee
where $a$ is some positive  number. 

For real negative $|\sigma|\ll 1$, we can write down a Lorentzian crossing equation. In particular, discussion of section \ref{section_correlators} implies that 
\be\label{app3}
G(\eta, \sigma)|_{\O_p}=(-1)^{\ell_p} \(G(\eta, |\sigma|)|_{\O_p}\)^*\ .
\ee
Hence, if we rotate sigma $\sigma \rightarrow |\sigma|e^{-i\pi}$ in equation (\ref{app2}), that must be consistent with the above relation. The contribution $G^{(p)}_{int}(\eta,\sigma)$ indeed satisfies this requirement. On the other hand, $G_0(\eta,\sigma)|_{\O_p}$ in general does not obey the crossing relation. This implies that $G(\eta,\sigma)|_{\O_p}$ must contain an imaginary part with non-integer powers of $\sigma$, which we have denoted as $G^{(p)}_{nint}(\eta,\sigma)$, such that the combination $G_0(\eta,\sigma)|_{\O_p}+G^{(p)}_{nint}(\eta,\sigma)$ has the right behavior. 

To be specific, let us consider a term $C^{(p)}_{n}\eta^{\frac{\tau_p}{2}} \sigma^{\Delta_p+n}  $ from the expansion (\ref{app1}). Clearly, this term is not consistent with the crossing equation (\ref{app3}). So, $G^{(p)}_{nint}(\eta,\sigma)$ must contain a similar term $i \tilde{C}^{(p)}_{n}\eta^{\frac{\tau_p}{2}} \sigma^{\Delta_p+n}$ such that $(C^{(p)}_{n}+i \tilde{C}^{(p)}_{n})\eta^{\frac{\tau_p}{2}} \sigma^{\Delta_p+n}$ is consistent with crossing. This imposes
\be
e^{-i\pi \Delta_p}\(C^{(p)}_{n}+i \tilde{C}^{(p)}_{n}\)=(-1)^{\ell_p}\(C^{(p)}_{n}-i \tilde{C}^{(p)}_{n}\)
\ee
implying\footnote{Note that $n$ is an even integer.} 
\be\label{app07}
\tilde{C}^{(p)}_{n}=C^{(p)}_{n} \tan \left(\frac{1}{2} \pi  (\Delta_p +\ell_p)\right)\ .
\ee
This relation is a manifestation of the fact that the lightcone limit conformal block (\ref{app2}) is valid even on the lower-half complex-$\sigma$ plane.\footnote{Clearly, the relation (\ref{app07}) blows up when $\Delta_p+\ell_p$ is an odd integer. In that case, we should  add terms like $i\sigma^{\Delta_p} \log \sigma$ in $G^{(p)}_{nint}(\eta,\sigma)$. Alternatively, we can treat $\Delta_p$ as a non-integer and take the integer limit at the end.}  So, $G(\eta,\sigma)$ in the Lorentzian lightcone limit for $\mbox{Im}\ \sigma<0$ can be expressed as an asymptotic series which is organized by twist
\be\label{app4}
G(\eta,\sigma) \sim 1 +\sum_{\tau_p\le \tau_{cutoff}} G(\eta,\sigma)|_{\O_p} \ ,
\ee
where we have isolated the identity contribution. Note that this discussion applies to all correction terms in $\delta G^{(p)}_{nint}(\eta,\sigma)$ as well.

Note that along the contour \ref{contour}
\be\label{appsk}
 \oint d\sigma\ \sigma^m  \sigma^{\Delta_p+n}=0  
\ee
for any non-negative integer $m$. Hence, we can define the following generalized correlator on the lower-half $\sigma$ plane
\be
G_0^{(- )}(\eta,\sigma)\sim 1+\sum_{p\neq \bold{1}}^{\tau_p\le \tau_{cutoff}} \(G_0(\eta,\sigma)|_{\O_p}+G^{(p)}_{nint}(\eta,\sigma)+\delta G^{(p)}_{nint}(\eta,\sigma)\)\ ,
\ee
where we have again isolated the identity contribution. Clearly, the generalized correlator has the following properties 
\be\label{bck2}
\mbox{Re}\ G_0^{(- )}(\eta,\sigma) = G_0(\eta,\sigma) \qquad \text{for} \qquad \mbox{Im}\ \sigma=0
\ee
and for any integer $m$
\be\label{bck3}
\oint d\sigma\ \sigma^m  G_0^{(- )}(\eta,\sigma)=0  
\ee
along the contour \ref{contour}. This now implies that 
\be
\mbox{Re} \oint d\sigma\ \sigma^m \(G_0^{(- )}(\eta,\sigma)-G(\eta,\sigma)\)=0
\ee
for any non-negative integer $m$. Therefore, we can write
\be\label{final_app}
\mbox{Re} \int_{-R}^R d\sigma\ \sigma^{m} \(G_0(\eta,\sigma)-G(\eta,\sigma)\)=\mbox{Re} \int_S d\sigma\ \sigma^{m} \sum_p G^{(p)}_{int}(\eta,\sigma)
\ee
where, $S=\{R e^{i\theta}, \theta\in [0,-\pi]\}$ with $0<\eta\ll R\ll 1$. To be precise, we should also include the correction term $\delta G^{(p)}_{int}(\eta,\sigma)$ in the right hand side. Since, however, terms in $\delta G^{(p)}_{int}(\eta,\sigma)$ are always subleading compared to terms in $ G^{(p)}_{int}(\eta,\sigma)$, for our purpose we can safely ignore $\delta G^{(p)}_{int}(\eta,\sigma)$. This immediately implies that we can use the identity (\ref{sachin}) to project to different powers of $1/\sigma$ obtaining the sum rule 
\be
\mbox{Re} \int_{-R}^R d\sigma\ \sigma^{\ell-2} \(G_0(\eta,\sigma)-G(\eta,\sigma)\)=\sum_{\ell'\ge \ell}C_{\ell', \ell}\ \eta^{\frac{\t_{\ell'}}{2}}\ , 
\ee
where, $ \ell\ge 2$ is an integer. 

It is clear from the derivation of the sum rule that terms in (\ref{bck1}) with positive powers of $\sigma$ do not contribute in the above sum rule. Any such term in (\ref{bck1}) can be absorbed in the definition of $G_0^{(- )}(\eta,\sigma)$ without affecting (\ref{bck2}) and (\ref{bck3}).

\subsection{Scalar Example}
For the purpose of demonstration, let us consider the special case where $X=\psi$ is a scalar primary. We can use the explicit lightcone conformal block derived by Dolan and Osborn in \cite{Dolan:2000ut} to obtain 
\be
G_0(\eta,\sigma)|_{\O_p}=a_p \eta^{\frac{\tau_p}{2} } \sigma ^{\Delta_p }\frac{\sqrt{\pi }\Gamma \left(\frac{\ell_p+\Delta_p }{2}\right) \, _2F_1\left(\frac{1}{2},\frac{\ell_p+\Delta_p }{2};\frac{1}{2} (\ell_p+\Delta_p +1); \sigma^2\right)}{2  \Gamma \left(\frac{1}{2} (\ell_p+\Delta_p +1)\right)}+\cdots\ ,
\ee
where terms that are suppressed by higher powers of $\eta$ are represented by dots and 
\be
a_p=2^{\Delta_p+\ell_p-2}\frac{(-1)^{\ell_p}    c_{\O\O^\dagger\O_p} c_{\psi\psi^\dagger\O_p} 2^{\Delta_p} \Gamma \left(\frac{\Delta_p+\ell_p+1}{2}\right)}{c_p \sqrt{\pi}\Gamma \left(\frac{\Delta_p+\ell_p}{2} \right)^2}\ .
\ee
Similarly, we can analytic continue the Dolan-Osborn block along the path shown in figure \ref{dt1}   to obtain $G(\eta,\sigma)|_{\O_p}$. In particular, using appendix B of \cite{Kundu:2019zsl}, at the leading order in the Lorentzian lightcone limit  we find 
\begin{align}
G(\eta,\sigma)|_{\O_p}&=G_0(\eta,\sigma)|_{\O_p}\(1+i \tan\pi \(\frac{\Delta_p+\ell_p}{2}\)\)\nonumber\\
&-a_p \frac{\eta^{\frac{\tau_p}{2} }}{ \sigma^{\ell_p-1}}\frac{i \pi ^{3/2} \, _2F_1\left(\frac{1}{2},\frac{1}{2} (-\ell_p-\Delta_p +2);\frac{1}{2} (-\ell_p-\Delta_p +3);\sigma^2\right)}{\cos \left(\frac{1}{2} \pi  (\Delta_p +\ell_p)\right) \Gamma \left(\frac{\ell_p+\Delta_p }{2}\right) \Gamma \left(\frac{1}{2} (-\ell_p-\Delta_p +3)\right)}
\end{align}
for complex $|\sigma|<1$ with $\mbox{Re}\ \sigma>0$. This is completely consistent with the preceding discussion.

\section{Mixed Correlators in the Lightcone Limit}\label{appendixB}
We can make a similar argument for the mixed correlator
 \be
 G_{mixed}(\eta,\sigma)=\langle   \O_2^\dagger(\mathbf{1})\O_1(\rr)\O_2(-\rr)\O_1^\dagger(-\mathbf{1})\rangle 
 \ee
 to show that terms that decay for small $\sigma$ can be safely ignored even in section \ref{section_mixed}. However, since $\O_1$ and $\O_2$ are scalar primaries, we can provide a more direct argument. Again we only consider the non-trivial case where the dimensions of the exchanged operators are non-integers. 

First, we start with a simpler Lorentzian correlator 
\be\label{app_mixed}
 \tilde{G}_{mixed}(\eta,\sigma)=\langle   \O_2^\dagger(\mathbf{1})\O_1(\rr)\O_1^\dagger(-\mathbf{1})\O_2(-\rr)\rangle
\ee
which can be determined by using the Euclidean OPE even for $\sigma<1$. The contribution of the $\O_1\O_2 \rightarrow\O_p\rightarrow \O_1^\dagger \O_2^\dagger$ conformal block to the correlator $ \tilde{G}_{mixed}(\eta,\sigma)$ can be computed by using the lightcone OPE (\ref{lope2}) or  the lightcone conformal block derived by Dolan and Osborn in \cite{Dolan:2000ut}. At the leading order in the lightcone limit, for real positive $\sigma<1$ we obtain
\begin{align}
 \tilde{G}_{mixed}&(\eta,\sigma)|_{\O_p}=\tilde{a}_p \eta^{\frac{\tau_p-\Delta_1-\Delta_2}{2}}\frac{\sigma^{\Delta_p}}{\(1+\sigma\)^{\ell_p+\Delta_p}}\nonumber\\
 & \times\ _2F_1\left(\frac{1}{2} \left(-\Delta _{12}+\ell_p+\Delta _p\right),\frac{1}{2} \left(\Delta _{12}+\ell_p+\Delta _p\right);\ell_p+\Delta _p;\frac{4 \sigma }{(\sigma +1)^2}\right)
\end{align}
where
\be
\tilde{a}_p=\(-\frac{1}{2}\)^{\ell_p}\frac{ c_{\O_1\O_2\O_p} c_{\O_1^\dagger\O_2^\dagger\O_p}}{2^{2(-\Delta_p+\Delta_1+\Delta_2)}c_p}\ .
\ee
Similarly, we can analytic continue the lightcone conformal block along the path shown in figure \ref{dt1}   to obtain  $\tilde{G}_{mixed}(\eta,\sigma)_{\O_p}$. In particular, using appendix B of \cite{Kundu:2019zsl}, at the leading order in the Lorentzian lightcone limit,   for real positive $\sigma< 1$,  we find (when $\Delta_p$ is not an integer)
\begin{align}\label{appmp}
 G_{mixed}&(\eta,\sigma)|_{\O_p}=\(1+i \frac{\cos \left(\pi  \Delta _{12}\right)-\cos  \left(\pi  \left(\ell_p+\Delta _p\right)\right)}{\sin \left(\pi  \left(\ell_p+\Delta _p\right)\right.}\)\tilde{G}_{mixed}(\eta,\sigma)|_{\O_p}\\
 &+\frac{8\pi i  \tilde{a}_p }{2^{2(\Delta_p+\ell_p)}} \frac{\eta^{\frac{\tau_p-\Delta_1-\Delta_2}{2}}}{\sigma^{\ell_p-1}}\frac{\(1+\sigma\)^{\ell_p+\Delta_p-2}\Gamma \left(\ell_p+\Delta _p-1\right) \Gamma \left(\ell_p+\Delta _p\right)}{\Gamma \left(\frac{1}{2} \left(\ell_p-\Delta _{12}+\Delta _p\right)\right)^2 \Gamma \left(\frac{1}{2} \left(\ell_p+\Delta _{12}+\Delta _p\right)\right)^2}\nonumber\\
 & \times\ _2F_1 \left(\frac{1}{2} \left(-\Delta _{12}-\ell_p-\Delta _p+2\right),\frac{1}{2} \left(\Delta _{12}-\ell_p-\Delta _p+2\right),-\ell_p-\Delta _p+2,\frac{4 \sigma }{(\sigma +1)^2}\right)\ .\nonumber
\end{align}
Clearly, terms with non-integer powers of $\sigma$ can only come from the first line. Similar to the previous case, we can utilize equation (\ref{appsk}) to analytically continue  (\ref{app_mixed}) to the lower-half $\sigma$ plane in the lightcone limit
\be
\tilde{G}_{mixed}^{(- )}(\eta,\sigma)\sim \sum_{p} \(1+i \frac{\cos \left(\pi  \Delta _{12}\right)-\cos  \left(\pi  \left(\ell_p+\Delta _p\right)\right)}{\sin \left(\pi  \left(\ell_p+\Delta _p\right)\right.}\)\tilde{G}_{mixed}(\eta,\sigma)|_{\O_p}
\ee
which enjoys the following properties. First, for any integer $m$
\be\label{bck4}
\oint d\sigma\ \sigma^m  \tilde{G}_{mixed}^{(- )}(\eta,\sigma)=0  
\ee
along the contour \ref{contour}. Secondly, for real $\sigma|<1$
\be\label{bck5}
\mbox{Re}\ \tilde{G}_{mixed}^{(- )}(\eta,\sigma)=\mbox{Re}\ \tilde{G}_{mixed}(\eta,\sigma)\ .
\ee
The last relation is rather obvious for positive $\sigma$. For negative $\sigma$, the above relation can be derived by using the lightcone conformal block. The above two relations are sufficient to conclude that terms in (\ref{appmp}) with non-integer powers of $\sigma$ do not contribute in the argument of section \ref{section_mixed}. Moreover, note that terms in (\ref{appmp}) with integer but positive powers of $\sigma$ can be absorbed in the definition of $\tilde{G}_{mixed}^{(- )}(\eta,\sigma)$ without affecting (\ref{bck4}) and (\ref{bck5}). Hence, only terms in (\ref{appmp}) that grow for small $\sigma$ contribute in the semicircle integral.

%%%%%%%%%%%%%%%%%%%%%%%%%%%%%%%%%%%%%%%%%%%

\end{spacing}


\begin{thebibliography}{999}

\bibitem{Wilson:1971bg} 
  K.~G.~Wilson,
  ``Renormalization group and critical phenomena. 1. Renormalization group and the Kadanoff scaling picture,''
  Phys.\ Rev.\ B {\bf 4}, 3174 (1971).
  doi:10.1103/PhysRevB.4.3174
  %%CITATION = doi:10.1103/PhysRevB.4.3174;%%
  
  %\cite{Wilson:1971dh}
\bibitem{Wilson:1971dh} 
  K.~G.~Wilson,
  ``Renormalization group and critical phenomena. 2. Phase space cell analysis of critical behavior,''
  Phys.\ Rev.\ B {\bf 4}, 3184 (1971).
  doi:10.1103/PhysRevB.4.3184
  %%CITATION = doi:10.1103/PhysRevB.4.3184;%%
  
  \bibitem{Zimmermann}
  W.~ Zimmermann,
  ``Lectures on Elementary Particles and Quantum Field Theory,"
   Brandeis Summer Institute in Theoretical Physics. MIT Press, Cambridge, Mass., 1970.


\bibitem{Mack:1976pa} 
  G.~Mack,
  ``Convergence of Operator Product Expansions on the Vacuum in Conformal Invariant Quantum Field Theory,''
  Commun.\ Math.\ Phys.\  {\bf 53}, 155 (1977).
  doi:10.1007/BF01609130
  %%CITATION = doi:10.1007/BF01609130;%%
  
  %\cite{Pappadopulo:2012jk}
\bibitem{Pappadopulo:2012jk} 
  D.~Pappadopulo, S.~Rychkov, J.~Espin and R.~Rattazzi,
  ``OPE Convergence in Conformal Field Theory,''
  Phys.\ Rev.\ D {\bf 86}, 105043 (2012)
  doi:10.1103/PhysRevD.86.105043
  [arXiv:1208.6449 [hep-th]].
  %%CITATION = doi:10.1103/PhysRevD.86.105043;%%


%\cite{Komargodski:2012ek}
\bibitem{Komargodski:2012ek} 
  Z.~Komargodski and A.~Zhiboedov,
  ``Convexity and Liberation at Large Spin,''
  JHEP {\bf 1311}, 140 (2013)
  doi:10.1007/JHEP11(2013)140
  [arXiv:1212.4103 [hep-th]].
  %%CITATION = doi:10.1007/JHEP11(2013)140;%%
  
%\cite{Nachtmann:1973mr}
\bibitem{Nachtmann:1973mr} 
  O.~Nachtmann,
  ``Positivity constraints for anomalous dimensions,''
  Nucl.\ Phys.\ B {\bf 63}, 237 (1973).
  doi:10.1016/0550-3213(73)90144-2
  %%CITATION = doi:10.1016/0550-3213(73)90144-2;%%
  
  
  %\cite{Costa:2017twz}
\bibitem{Costa:2017twz} 
  M.~S.~Costa, T.~Hansen and J.~Penedones,
  ``Bounds for OPE coefficients on the Regge trajectory,''
  JHEP {\bf 1710}, 197 (2017)
  doi:10.1007/JHEP10(2017)197
  [arXiv:1707.07689 [hep-th]].
  %%CITATION = doi:10.1007/JHEP10(2017)197;%%

 %\cite{Caron-Huot:2017vep}
\bibitem{Caron-Huot:2017vep} 
  S.~Caron-Huot,
  ``Analyticity in Spin in Conformal Theories,''
  JHEP {\bf 1709}, 078 (2017)
  doi:10.1007/JHEP09(2017)078
  [arXiv:1703.00278 [hep-th]].
  %%CITATION = doi:10.1007/JHEP09(2017)078;%%
  
  %\cite{Hartman:2016lgu}
\bibitem{Hartman:2016lgu} 
  T.~Hartman, S.~Kundu and A.~Tajdini,
  ``Averaged Null Energy Condition from Causality,''
  JHEP {\bf 1707}, 066 (2017)
  doi:10.1007/JHEP07(2017)066
  [arXiv:1610.05308 [hep-th]].
  %%CITATION = doi:10.1007/JHEP07(2017)066;%%

%\cite{Fitzpatrick:2012yx}
\bibitem{Fitzpatrick:2012yx} 
  A.~L.~Fitzpatrick, J.~Kaplan, D.~Poland and D.~Simmons-Duffin,
  ``The Analytic Bootstrap and AdS Superhorizon Locality,''
  JHEP {\bf 1312}, 004 (2013)
  doi:10.1007/JHEP12(2013)004
  [arXiv:1212.3616 [hep-th]].
  %%CITATION = doi:10.1007/JHEP12(2013)004;%%
  
%\cite{Simmons-Duffin:2016wlq}
\bibitem{Simmons-Duffin:2016wlq} 
  D.~Simmons-Duffin,
  ``The Lightcone Bootstrap and the Spectrum of the 3d Ising CFT,''
  JHEP {\bf 1703}, 086 (2017)
  doi:10.1007/JHEP03(2017)086
  [arXiv:1612.08471 [hep-th]].
  %%CITATION = doi:10.1007/JHEP03(2017)086;%%

%\cite{Alday:2015eya,Alday:2015ota,Alday:2016njk}
\bibitem{Alday:2015eya}
L.~F.~Alday, A.~Bissi and T.~Lukowski,
``Large spin systematics in CFT,''
JHEP \textbf{11}, 101 (2015)
doi:10.1007/JHEP11(2015)101
[arXiv:1502.07707 [hep-th]].

%\cite{Alday:2015ota}
\bibitem{Alday:2015ota}
L.~F.~Alday and A.~Zhiboedov,
``Conformal Bootstrap With Slightly Broken Higher Spin Symmetry,''
JHEP \textbf{06}, 091 (2016)
doi:10.1007/JHEP06(2016)091
[arXiv:1506.04659 [hep-th]].

%\cite{Alday:2016njk}
\bibitem{Alday:2016njk}
L.~F.~Alday,
``Large Spin Perturbation Theory for Conformal Field Theories,''
Phys. Rev. Lett. \textbf{119}, no.11, 111601 (2017)
doi:10.1103/PhysRevLett.119.111601
[arXiv:1611.01500 [hep-th]].



  %\cite{Osterwalder:1973dx}
\bibitem{Osterwalder:1973dx} 
  K.~Osterwalder and R.~Schrader,
  ``Axioms For Euclidean Green's Functions,''
  Commun.\ Math.\ Phys.\  {\bf 31}, 83 (1973).
  doi:10.1007/BF01645738
  %%CITATION = doi:10.1007/BF01645738;%%
  
  %\cite{Osterwalder:1974tc}
\bibitem{Osterwalder:1974tc} 
  K.~Osterwalder and R.~Schrader,
  ``Axioms for Euclidean Green's Functions. 2.,''
  Commun.\ Math.\ Phys.\  {\bf 42}, 281 (1975).
  doi:10.1007/BF01608978
  %%CITATION = doi:10.1007/BF01608978;%%
  
  \bibitem{slava1}
  P.~Kravchuk, J.~Qiao and S.~Rychkov,
  ``Distributions in CFT I. Cross-Ratio Space,''
  arXiv:2001.08778 [hep-th].
  %%CITATION = ARXIV:2001.08778;%%
 
 \bibitem{slava2}
  P.~ Kravchuk, J.~ Qiao, S.~ Rychkov,
 ``Distributions in CFT II. Minkowski Space,"
 and
 ``Distributions in CFT III. Lorentzian Cylinder,"
Work in progress.

\bibitem{haag} 
  R.~Haag,
  \textit{Local quantum physics: Fields, particles, algebras},
  Berlin, Germany: Springer (1992). 
  R.~F.~Streater and A.~S.~Wightman,
  \textit{PCT, spin and statistics, and all that,}
  New York, USA: WA Benjamin (1964).  

  
 
   
  
  %\cite{Casini:2010bf}
\bibitem{Casini:2010bf} 
  H.~Casini,
  ``Wedge reflection positivity,''
  J.\ Phys.\ A {\bf 44}, 435202 (2011)
  doi:10.1088/1751-8113/44/43/435202
  [arXiv:1009.3832 [hep-th]].
  %%CITATION = doi:10.1088/1751-8113/44/43/435202;%%
  
  %\cite{Maldacena:2015waa}
\bibitem{Maldacena:2015waa} 
  J.~Maldacena, S.~H.~Shenker and D.~Stanford,
  ``A bound on chaos,''
  JHEP {\bf 1608}, 106 (2016)
  doi:10.1007/JHEP08(2016)106
  [arXiv:1503.01409 [hep-th]].
  %%CITATION = doi:10.1007/JHEP08(2016)106;%%
  
 %\cite{Rosso:2019txh}
\bibitem{Rosso:2019txh}
F.~Rosso,
``Global aspects of conformal symmetry and the ANEC in dS and AdS,''
JHEP \textbf{03}, 186 (2020)
doi:10.1007/JHEP03(2020)186
[arXiv:1912.08897 [hep-th]].
 
  
  %\cite{Hartman:2015lfa}
\bibitem{Hartman:2015lfa} 
  T.~Hartman, S.~Jain and S.~Kundu,
  ``Causality Constraints in Conformal Field Theory,''
  JHEP {\bf 1605}, 099 (2016)
  doi:10.1007/JHEP05(2016)099
  [arXiv:1509.00014 [hep-th]].
  %%CITATION = doi:10.1007/JHEP05(2016)099;%%

  

  \bibitem{Li:2015rfa} 
  D.~Li, D.~Meltzer and D.~Poland,
  ``Non-Abelian Binding Energies from the Lightcone Bootstrap,''
  JHEP {\bf 1602}, 149 (2016)
  doi:10.1007/JHEP02(2016)149
  [arXiv:1510.07044 [hep-th]].
  %%CITATION = doi:10.1007/JHEP02(2016)149;%%
  
%\cite{Costa:2011mg}
\bibitem{Costa:2011mg} 
  M.~S.~Costa, J.~Penedones, D.~Poland and S.~Rychkov,
  ``Spinning Conformal Correlators,''
  JHEP {\bf 1111}, 071 (2011)
  doi:10.1007/JHEP11(2011)071
  [arXiv:1107.3554 [hep-th]].
  %%CITATION = doi:10.1007/JHEP11(2011)071;%%



%\cite{Dolan:2000ut}
\bibitem{Dolan:2000ut} 
  F.~A.~Dolan and H.~Osborn,
  ``Conformal four point functions and the operator product expansion,''
  Nucl.\ Phys.\ B {\bf 599}, 459 (2001)
  doi:10.1016/S0550-3213(01)00013-X
  [hep-th/0011040].
  %%CITATION = doi:10.1016/S0550-3213(01)00013-X;%%


%\cite{ElShowk:2012ht}
\bibitem{ElShowk:2012ht} 
  S.~El-Showk, M.~F.~Paulos, D.~Poland, S.~Rychkov, D.~Simmons-Duffin and A.~Vichi,
  ``Solving the 3D Ising Model with the Conformal Bootstrap,''
  Phys.\ Rev.\ D {\bf 86}, 025022 (2012)
  doi:10.1103/PhysRevD.86.025022
  [arXiv:1203.6064 [hep-th]].
  %%CITATION = doi:10.1103/PhysRevD.86.025022;%%

%\cite{El-Showk:2014dwa}
\bibitem{El-Showk:2014dwa} 
  S.~El-Showk, M.~F.~Paulos, D.~Poland, S.~Rychkov, D.~Simmons-Duffin and A.~Vichi,
  ``Solving the 3d Ising Model with the Conformal Bootstrap II. c-Minimization and Precise Critical Exponents,''
  J.\ Stat.\ Phys.\  {\bf 157}, 869 (2014)
  doi:10.1007/s10955-014-1042-7
  [arXiv:1403.4545 [hep-th]].
  %%CITATION = doi:10.1007/s10955-014-1042-7;%%
  
  
  
  %\cite{Kos:2014bka}
\bibitem{Kos:2014bka} 
  F.~Kos, D.~Poland and D.~Simmons-Duffin,
  ``Bootstrapping Mixed Correlators in the 3D Ising Model,''
  JHEP {\bf 1411}, 109 (2014)
  doi:10.1007/JHEP11(2014)109
  [arXiv:1406.4858 [hep-th]].
  %%CITATION = doi:10.1007/JHEP11(2014)109;%%
  
  %\cite{Simmons-Duffin:2015qma}
\bibitem{Simmons-Duffin:2015qma} 
  D.~Simmons-Duffin,
  ``A Semidefinite Program Solver for the Conformal Bootstrap,''
  JHEP {\bf 1506}, 174 (2015)
  doi:10.1007/JHEP06(2015)174
  [arXiv:1502.02033 [hep-th]].
  %%CITATION = doi:10.1007/JHEP06(2015)174;%%

%\cite{Kos:2016ysd}
\bibitem{Kos:2016ysd} 
  F.~Kos, D.~Poland, D.~Simmons-Duffin and A.~Vichi,
  ``Precision Islands in the Ising and $O(N)$ Models,''
  JHEP {\bf 1608}, 036 (2016)
  doi:10.1007/JHEP08(2016)036
  [arXiv:1603.04436 [hep-th]].
  %%CITATION = doi:10.1007/JHEP08(2016)036;%%




 %\cite{Brower:2006ea}
\bibitem{Brower:2006ea} 
  R.~C.~Brower, J.~Polchinski, M.~J.~Strassler and C.~I.~Tan,
  ``The Pomeron and gauge/string duality,''
  JHEP {\bf 0712}, 005 (2007)
  doi:10.1088/1126-6708/2007/12/005
  [hep-th/0603115].
  %%CITATION = doi:10.1088/1126-6708/2007/12/005;%%
  
  
%\cite{Cornalba:2007fs}
\bibitem{Cornalba:2007fs} 
  L.~Cornalba,
  ``Eikonal methods in AdS/CFT: Regge theory and multi-reggeon exchange,''
  arXiv:0710.5480 [hep-th].
  %%CITATION = ARXIV:0710.5480;%%

%\cite{Cornalba:2008qf}
\bibitem{Cornalba:2008qf} 
  L.~Cornalba, M.~S.~Costa and J.~Penedones,
  ``Eikonal Methods in AdS/CFT: BFKL Pomeron at Weak Coupling,''
  JHEP {\bf 0806}, 048 (2008)
  doi:10.1088/1126-6708/2008/06/048
  [arXiv:0801.3002 [hep-th]].
  %%CITATION = doi:10.1088/1126-6708/2008/06/048;%%
  
  %\cite{Costa:2012cb}
\bibitem{Costa:2012cb} 
  M.~S.~Costa, V.~Goncalves and J.~Penedones,
  ``Conformal Regge theory,''
  JHEP {\bf 1212}, 091 (2012)
  doi:10.1007/JHEP12(2012)091
  [arXiv:1209.4355 [hep-th]].
  %%CITATION = doi:10.1007/JHEP12(2012)091;%%

%\cite{Afkhami-Jeddi:2016ntf}
\bibitem{Afkhami-Jeddi:2016ntf} 
  N.~Afkhami-Jeddi, T.~Hartman, S.~Kundu and A.~Tajdini,
  ``Einstein gravity 3-point functions from conformal field theory,''
  JHEP {\bf 1712}, 049 (2017)
  doi:10.1007/JHEP12(2017)049
  [arXiv:1610.09378 [hep-th]].
  %%CITATION = doi:10.1007/JHEP12(2017)049;%%

%\cite{Kundu:2019zsl}
\bibitem{Kundu:2019zsl} 
  S.~Kundu,
  ``Renormalization Group Flows, the $a$-Theorem and Conformal Bootstrap,''
  arXiv:1912.09479 [hep-th].
  %%CITATION = ARXIV:1912.09479;%%
  
\end{thebibliography}
\end{document}